\RequirePackage{lineno}
\documentclass[twocolumn,showpacs,preprintnumbers,amsmath,amssymb,superscriptaddress]{revtex4}

\usepackage{graphicx}
\usepackage{dcolumn}
\usepackage{bm}
\usepackage{colordvi}
\usepackage{color}
\usepackage{epsfig}
\usepackage{graphicx}
\usepackage{color}

\DeclareMathOperator{\sgn}{sgn}

\let\vec\bm

\def\coloronline{(Color online)\ }

\begin{document}
\title{Structural relaxation of polydisperse hard spheres: comparison of the mode-coupling theory to a Langevin dynamics simulation}

\def\unikn{\affiliation{
Fachbereich Physik, Universit\"at Konstanz, 78457 Konstanz, Germany
}}
\def\almeria{\affiliation{
Departamento de F\'{\i}sica Aplicada , Universidad de Almer\'{\i}a, 04120 Almer\'{\i}a, Spain
}}
\def\zk{\affiliation{
Zukunftskolleg, Universit\"at Konstanz, 78457 Konstanz, Germany
}}
\def\dlr{\affiliation{
Institut f\"ur Materialphysik im Weltraum, Deutsches Zentrum f\"ur Luft- und Raumfahrt (DLR), 51170 K\"oln, Germany
}}
\author{F.~Weysser}\unikn
\author{A.~M.~Puertas}\almeria
\author{M.~Fuchs}\unikn
\author{Th.~Voigtmann}\unikn\dlr\zk


\begin{abstract}
We analyze the slow, glassy structural relaxation as measured through
collective and tagged-particle
density correlation functions obtained from Brownian
dynamics simulations for a polydisperse system of quasi-hard spheres
in the framework of the mode-coupling theory of the glass transition (MCT).
Asymptotic analyses show good agreement for the collective dynamics when
polydispersity effects are taken into account in a multi-component calculation, 
but qualitative disagreement at small $q$ when the system is treated as effectively
monodisperse. The origin of the different small-$q$ behaviour is attributed
to the interplay between interdiffusion processes and structural
relaxation. Numerical solutions of the MCT equations are obtained
taking properly binned partial static structure factors from the simulations
as input. Accounting for a shift in the critical density, the collective
density correlation functions are well described by the theory at all
densities investigated in the simulations, with quantitative agreement
best around the maxima of the static structure factor, and worst around its
minima. A parameter-free comparison of the tagged-particle dynamics
however reveals large quantiative errors for small wave numbers that are
connected to the well-known decoupling of self-diffusion from structural
relaxation and to dynamical heterogeneities. While deviations from MCT
behaviour are clearly seen in the tagged-particle quantities for densities
close to and on the liquid side of the MCT glass transition, no such
deviations are seen in the collective dynamics.
\end{abstract}

\maketitle 
\section{Introduction}

Understanding the slow dynamical processes occurring in supercooled
glass-forming liquids is still one of the challenges in condensed matter
physics. The mode-coupling theory of the glass transition (MCT), introduced
in 1984 by Bengtzelius, G\"otze, and Sj\"olander and Leutheusser
\cite{BengtMCT,Leuth}, provides a quantitative description of the initial
slowing down of structural relaxation, when approaching the glassy state
from the liquid region. The theory predicts an ideal glass transition
whose signature is a two-step relaxation of dynamical density correlation
functions. It arises from a divergence of two time scales connected with
the intermediate relaxation ascribed to relaxation of particles inside their
neighbor-cages (the $\beta$ relaxation), and with the final escape of
particles from their initial positions that restores liquid-like motion
(the $\alpha$ relaxation process). Two hallmarks of glassy dynamics,
viz.\ nonexponential, ``stretched-exponential''
$\alpha$ relaxation and its scaling (also known as
``time-temperature superposition principle''), are predicted as asymptotic
results of MCT. The theory stimulated many experiments specifically
to address the dynamical window for which the theory was designed;
among them dynamic light scattering performed on colloidal systems,
Brillouin and neutron scattering, dielectric spectroscopy, and computer
simulation studies (see Ref.~\cite{goetze_cond,goetzeBuch,kob1999,KobBinderBuch}
for reviews).

However, MCT is based on the ad-hoc assumption that the fluctuating forces
(the longitudinal projections of the microscopic stresses for a particular
wave vector $\vec q$) are governed entirely by the dynamics of density
pair fluctuations. To close the equations, one then further approximates
a dynamical four-point average through a product of density correlation
functions. Even though the theory has had many successes in
describing some key features of the slow dynamics, often validated through
comparison of its asymptotic formul\ae\ with experimental and simulation
data, the accuracy of the MCT approximation is still largely unknown.
In molecular glass formers, the fact that relaxation times do not diverge
at the MCT transition, but continue to grow smoothly in a regime where
motion is thought to be no longer liquid-like but governed by activated,
so-called ``hopping'', processes, is the most widely criticized
manifestation of the approximate nature of MCT.
Another commonly quoted feature that is not contained in the theory are
the non-Gaussian distributions of particle displacement discussed in terms
of dynamical heterogeneity \cite{kob1997, glotzer2002}. This effect is
often linked to the appearance
of a decoupling of viscous and diffusive time scales -- the
breakdown of the Stokes-Einstein relation, although it could be argued that
not its breakdown at low temperatures, but rather its validity at higher temperatures in
complex glass formers is the surprising feature.

If hopping processes are indeed what is missing in MCT, checking the
feasibility of creating an atomistic model system where such effects are
absent, is an obvious thing to do. Following the pioneering dynamic
light-scattering experiments on colloidal hard-sphere-like suspensions by
Pusey, van~Megen and coworkers \cite{Pusey_vanMegen1986, Pusey_vanMegen1987, Pusey1989, vanMegen_Underwood1994, vanMegen_Mortenson1998}, 
hard spheres with Brownian short-time motion have sometimes been quoted
in this regard. Yet, this exceptional nature of the hard-sphere glass
transition has been challenged based on computer simulation of tagged-particle
density correlation functions \cite{FlennerSzamel2005}. In this contribution, we wish
to analyze the situation further by shifting focus from the incoherent
quantities to the collective density correlation functions that are
closer to the framework of MCT. We find that, a number of commonly
discussed shortcomings of the theory appears only in the
tagged-particle but not in the collective dynamics.

One strength of MCT is that it allows, in principle, to predict
detailed information on the slow dynamics when given only the particles'
interaction potentials, in the form of the static structure factor, as
input. For real-world glass formers, generally mixtures or moderately
complicated organic molecules, resolving all the partial static
structure information required is a formidable task. Even more so if one is
interested
not only in the static structure, but also the corresponding dynamical
relaxation functions. Thus, in many experimental
studies, MCT results were taken either
from asymptotic expansions (that cannot address the molecular details and
preasymptotic corrections, which may be strong), or from schematic
simplifications of the theory's equations (resulting in a set of fit
parameters whose physical meaning is rather unclear).
Only recently has it become possible to perform MCT calculations
based on actual experimentally measured partial structure factors,
due to advances in neutron scattering techniques on liquid metallic melts
\cite{voigtmann_euro}.

Thus, testing the ``full MCT'',
that is, putting to test the dynamics as calculated within the theory from
the static structure factor (without invoking asymptotic or schematic
limits of the theory's equations) against the measured one,
is a task for molecular dynamics simulations, and has been performed on
the standard glass-forming
binary Lennard-Jones mixture \cite{nauroth1997,kob2002,
FlennerSzamel2005, FlennerSzamel2005_2}, on hard-sphere mixtures \cite{foffi2003,foffi2004},
soft spheres with short-ranged attraction \cite{henrich2007, zaccarelli2006},
and in more complicated systems such as network-forming strong liquids
\cite{sciortino2001,voigtmann_horbach}, metallic glasses \cite{teichler2001},
polymer melts \cite{chong2007}, or computer models
of organic glass formers such as ortho-terphenyl \cite{rinaldi2001,chong2004}.
As it turns out, these systems are already quite demanding for MCT,
although the theory fares well in a qualitative description of the dynamical
phenomena, sometimes even quantiatively (most notably,
Ref.~\cite{sciortino2001}, where also static triplet correlation functions
have been extracted from simulation and fed into MCT, addressing a term
in the MCT equations whose existence is often silently ignored).

The most simple model for a classical dense liquid is arguably the hard
sphere system.
In a previous study, we addressed a test of MCT for this system
partially, by comparing
MCT and molecular dynamics simulations for a polydisperse quasi-hard
sphere system \cite{voigtmann2004}. There, however, computational limitations
in acquiring the desired statistics restricted the discussion to the
single-particle dynamics (in form of the
incoherent density correlation functions and quantities derived from it,
such as the mean-squared displacements or diffusion coefficients).
It should be
stressed that MCT is, in its very essence, a theory for the \emph{collective}
slowing down caused by a feedback mechanism for the collective, or coherent
density correlation
functions. Calculating tagged-particle dynamics from this viewpoint involves
an additional level of (MCT-approximate) equations, and can thus be viewed
as a more indirect way of testing the theory.
In this paper, we complete the task of Ref.~\cite{voigtmann2004} by
detailing a comprehensive, quantitative comparison of MCT with molecular
dynamics computer simulations for the same quasi hard-sphere system
on the level of the collective density
correlation functions. Additionally, while in Ref.~\cite{voigtmann2004}
an approximate liquid-state theory for the static structure factor input
to MCT was used (the Percus-Yevick approximation), we avoid this additional
non-MCT level of approximations by using directly the simulated static
structure factors.

Comparing the dynamics to MCT, it should be
recognized that the theory focuses on the slow structural relaxation,
mistreating the short-time dynamics as governed by uncorrelated binary
collisions (whose inclusing into the theory is not straightforward
\cite{nauroth1997,kob2002}). It is thus desirable to minimize the influence
of this short-time dynamics, in particular since its details do not
change those of the long-time relaxation \cite{gleim2000, szamel2004};
it turns out that this is achieved by including
stochastic noise in the simulated equations of motion, leading to a
Langevin-dynamics simulation. Standard molecular-dynamics integration
is then most easily implemented using a regular soft-sphere potential,
$V(r)\propto r^{-36}$. For such steep power-law potentials, it is known
that the influence of slight `softness' on the dynamics can be mapped to
an effective density in a mapping that takes into account the according
shift of the freezing point \cite{Lange2009}.
In addition, the stochastic dynamics provides a link to experimental
data on colloidal hard-sphere-like systems, whose Brownian short-time
dynamics we mimick in our simulations.

Monodisperse hard- or soft-sphere systems beyond the freezing point readily crystallize in
simulation. To avoid this, we take
the spheres' radii to be polydisperse, evenly sampled from a narrow
distribution just wide enough to suppress crystallization on the time
scales considered in our simulations. It is well known that already small
polydispersities are very efficient in slowing down nucleation events
dramatically \cite{williams2008,zaccarelli2009}; polydispersity is also
inherent to most colloidal suspensions, making it a natural feature to
consider. Binary mixtures are another common
way of circumventing unwanted crystallization, but polydisperse systems
have the advantage that one can meaningfully construct species-averaged total
correlation functions that are not too different from the individual
partial correlations. In principle, this allows to greatly simplify the
discussion, by applying the original one-component formulation of MCT.
However, polydispersity may play an interesting role when comparing
theory and simulations in particular at small wave numbers $q$. We will
discuss these points in detail below, including three- and five-component
moment approximations to the polydisperse radius distribution in
multi-component MCT \cite{goetze1987b,voigtmann2003}.
Comparing with the one-component MCT, this addresses the question of
static versus dynamic averaging: while the true dynamics of the system
can be mapped onto an effective one-component system only by averaging
at the level of the dynamical correlation functions (a procedure to which
we will refer as post-averaging), it is tempting
to perform such averaging over a narrow polydispersity distribution
already on the level of the static structure factor (pre-averaging).
However, as we will
discuss, the nonlinear feedback effects of the dynamics pose a limit to
the validity of such an approach.

The paper is
organized as follows: In Section~\ref{basic} the features needed in the
further analysis of both MCT and simulation are reported for reference.
Section~\ref{asymptotic} demonstrates purely asymptotic analyses of the
simulation data, while Sec.~\ref{mctanalysis} turns to the full MCT
description of the simulated dynamical correlation functions.
Finally, in Sec.~\ref{concl} we summarize.

\section{Simulation and MCT}\label{basic}

\subsection{Molecular-dynamics simulation} \label{antoniotext}

We perform strongly damped molecular-dynamics (Langevin-dynamics) simulations
mimicking
colloidal Brownian dynamics (BD). The core-core respulsion between particles
$i$ and $j$ is given by
\begin{equation}
V_{ij}(r)=k_{\text{B}}T\left( \frac{r}{d_{ij}} \right)^{-36}\,,
\end{equation}
where $d_{12}$ is the center-to-center distance, $d_{ij}=(d_i+d_j)/2$, with
$d_i$ the diameters of the particles, sampled from a uniform distribution
centered on the mean diameter $d$ with half-width $\delta=0.1d$.
Such a soft-sphere system
has only one control parameter given by a specific combination of number
density $\rho$ and temperature $T$, $\Gamma=\rho T^{-12}$ $[\text{m}^{-3}
\text{K}^{-12}]$ \cite{mc_donald}. We vary $\Gamma$ by keeping the temperature
fixed and changing the system's density. It has been shown \cite{Lange2009}
that the exponent $n=36$ of the inverse power-law potential is large enough
to effectively approximate hard-sphere behavior.

The equation of motion for particle $j$ is given by the Langevin equation,
\begin{equation}\label{bdeq}
m \ddot{{\vec r}}_j = \sum_i {\vec F}_{ij}-\gamma\dot{\vec r}_j+{\vec\xi}_j(t)
\end{equation}
which contains the direct forces between particles ${\vec F}_{ij}$, and
stochastic and friction forces, with friction coefficient $\gamma$, modeling
interaction with a solvent. Assuming Stokes friction, its value would be
connected to the solvent viscosity $\eta_s$,
$\gamma = 3\pi d \eta_s$, where $d$ is the hydrodynamic diameter of the
particle (approximated as equal for all particles, since the spread in the
$d_i$ is small). The random forces fulfill the fluctuation-dissipation theorem,
$\langle {\vec\xi}_i(t) {\vec\xi}_j(t') \rangle=6 k_{\text{B}}T \gamma
\delta(t-t') \delta_{ij}$. Let us note that with the value of $\gamma$
chosen in our simulations, the short-time dynamics visible in
the correlators and in the mean-squared displacement is not yet completely
overdamped, i.e., it is not strictly
diffusive, but rather strongly damped ballistic. Since it is not our aim to
investigate the very short-time dynamical features of the simulations, this
will not be discussed in the following.

Equilibration runs were performed with undamped Newtonian dynamics in all
cases, since the damping introduces a slowing down in the overall time scale.
$N=1000$ particles are simulated in a cubic box with standard periodic boundary
conditions. Lengths are measured in units of the mean diameter $d$, the
particle mass $m=1$, and the unit of time is fixed setting the thermal velocity
to $v_{\text{th}}=({k_{\text{B}}T/m})^{1/2}=1/\sqrt{3}$. The damping was chosen as
$\gamma=20$ in these dimensionless units, and the equations of motion were
integrated following Heun's algorithm \cite{Paul1995} with a time step of
$\delta t=0.0005$.
Due to the polydispersity, crystallization did not occur in the runs
that have been analyzed for the following discussions. Crystallization
was monitored through the orientational order parameter $Q_6$
\cite{Steinhardt83,Wolde96}. It was found that $8$ out of $18$ runs for
$\varphi=0.58$, and $16$ out of $26$ runs for $\varphi=0.585$ in fact did
crystallize and
had to be excluded. Density is reported as volume fraction,
$\varphi=(\pi/6)d^3\left[1+\delta^2\right]\rho$, where the polydispersity
has been taken into account. Volume fractions investigated in the following
are $\varphi=0.50$, $0.53$, $0.55$, $0.57$, $0.58$, and $0.585$.
In all states $5$ or $10$ different systems were prepared from scratch,
and $100$ or $50$ independent correlation functions were measured adding to a
total statistics of $500$ evaluations of the dynamical quantities per state.

In order to study the specific effects of polydispersity, a varying number
of bins $M$ has been used to group 
particles according to their size: besides the usual
effective one-component analysis, $M=1$, we also discuss a three-component,
$M=3$, and a five-component, $M=5$, interpretation of the data. In these
cases, bins of uniform width have been chosen.
We mainly discuss the collective density correlation functions
(intermediate scattering functions),
\begin{equation}\label{correlator_definition}
\Phi_{\alpha \beta}(q,t)
  = \langle \varrho_\alpha(\vec q,t)^* \varrho_\beta(\vec q) \rangle\,,
\end{equation}
whose equal-time values are the partial
static structure factors
$S_{\alpha \beta}(q)= \Phi_{\alpha \beta}(q,t=0)$.
Here, indices $\alpha,\beta=1,\ldots M$ label the component bins
containing $N_{\alpha,\beta}$ particles, and
$\varrho_\alpha(\vec {q},t)= N^{-1/2}\sum_{k=1}^{N_\alpha}
\exp[i {\vec q} \cdot {\vec r}_\alpha^k(t)]$
are the collective partial-number-density fluctuations at wave vector
$\vec q$, where
$\vec{r}_\alpha^k(t)$ is the position of the k-th particle of species $\alpha$.
Note that these correlation functions are real-valued and depend on $\vec q$
only through its scalar invariant $q$, as the system is isotropic and
translational invariant.
The brackets indicate time-origin or canonical averaging in the
simulation or theoretical approach, respectively. To improve the statistics of
this function, a small dispersion in $q$-modulus was allowed for,
$\delta_q d=0.2/\sqrt{q}$ for $qd>6.0$ and $\delta_q d=0.2/\sqrt{3}$ for
$qd<6.0$. The change in $\delta_q$ can be noted as a kink in some wave-vector dependent quantities
derived from it at $qd=6.0$. 

Additionally, we measure the tagged-particle
(incoherent) density correlation functions, i.e., the self-intermediate
scattering function
\begin{equation}
\Phi^s_\alpha (q,t)
  = \langle \varrho_\alpha^s(\vec{q},t) \varrho_\alpha^s(\vec{q}) \rangle\,,
\end{equation}
where $\varrho_\alpha^s(\vec{q},t)= \exp[i \vec{q}\cdot\vec{r}_\alpha^{s}(t)]$
denotes the
Fourier transformed one-particle density. In this case, averaging over all
particles of the same species allows to further improve statistics in the
simulation. $\Phi^s_\alpha(q,t)$ is connected in the low-$q$ limit to
the mean squared displacement of species $\alpha$,
\begin{equation}
\delta r^2_\alpha(t)=
  \langle [\vec{r}_\alpha(t) - \vec{r}_\alpha(0) ]^2\rangle\,.
\end{equation}

The ``polydispersity-averaged'' total correlation functions are recovered
by summing over the bins,
\begin{equation}\label{summedphi}
\phi(q,t) = \frac{ \sum_{\alpha \beta} \Phi_{\alpha \beta}(q,t)}
  {\sum_{\alpha \beta} S_{\alpha \beta}(q)}
  = \frac{ \sum_{\alpha \beta} \Phi_{\alpha \beta}(q,t)}{S(q)}\,,
\end{equation}
which in the tagged-particle correlation function reduces to
\begin{equation}
\phi^s(q,t) = (1/M)\sum_{\alpha} \Phi_\alpha^s(q,t)
\end{equation}
and the analog expression for the mean-squared displacement,
\begin{equation}
\delta r^2 = 1/M  \sum_{\alpha} \delta r^2_\alpha(q,t)\,.
\end{equation}

\subsection{Mode-coupling theory}

The mode-coupling theory for $M$-component mixtures
builds upon an exact equation of motion derived
for the matrix of the partial dynamical density correlation functions
$\Phi_{\alpha\beta}(q,t)$ defined in Eq.~\eqref{correlator_definition}.
Applying Zwanzig-Mori projection operators to the Liouville equation
governing the microscopic dynamics, one arrives at \cite{goetze1987b}
\begin{multline} 
 \ddot{\bm{\Phi}}(q,t) + \bm{J}(q) \bm{S}^{-1}(q) \bm{\Phi}(q,t) 
\\
+ \bm{J}(q) \int \limits_{0}^{t} dt' \; \bm{M}(q, t-t') \dot{\bm{\Phi}}(q,t')=0\,.
\label{timespace_final}
\end{multline}
with $J_{\alpha \beta}(q)=q^2 \delta_{\alpha \beta} x_\alpha v^2_{\text{th},\alpha}$
and the memory kernel matrix $\bm{M}_q(t)$ which embodies the fluctuating
quantities and plays the role of a generalized friction coefficient.
Note that in our case, $v_{\text{th},\alpha}$ is the same for all
pseudo-species.
Initial
conditions for the equations of motion are $\bm{\Phi}(q,t=0)=:\bm{S}(q)$
and $ \dot{\bm{\Phi}}(q,t=0)=0$. We split the memory kernel into a contribution
describing the regular part of the friction, modeled as a Markov process
with a damping coefficient $\nu=\gamma$ that is chosen in agreement with the
one taken in the simulations, and a collective part
describing the slow dynamics,
\begin{equation}
 \bm{M}(q,t) \approx \bm{J}^{-1}(q) \bm{\nu}(q) \delta(t)+\bm{M}^{MCT}(q,t)\,,
\end{equation}
where $\nu_{\alpha\beta}(q)=\nu/M$  with $\nu = \gamma/2$. 
MCT now approximates $\bm{M}^{MCT}$ as a nonlinear functional of the
density correlation functions,
\begin{equation}
\bm{M}^{MCT}(q,t) = \mathcal{F}\left[ \bm{\Phi}(q,t)\right] 
\end{equation}
with components
\begin{multline}
\mathcal{F}_{\alpha \beta}\left[\bm{\Phi}(q,t) \right] =  \frac{1}{2q} \frac{\rho}{x_\alpha x_\beta} \sum \limits_{ \stackrel{\alpha^{''} \beta^{''}} {\alpha' \beta'}} \int \frac{d^3 k}{(2 \pi)^3} \mathcal{V}_{\alpha \alpha' \alpha''}(\vec{q},\vec{k},\vec{p})
\\ \times
  \Phi_{\alpha' \beta'}(k,t) \Phi_{\alpha'' \beta''}(p,t)  \mathcal{V}_{\beta \beta' \beta''}(\vec{q},\vec{k},\vec{p})\,.
\label{memorykernel}
\end{multline}
The vertices couple the density fluctuations of different modes, with $\vec{q}=\vec{k}+\vec{p}$,
\begin{equation}
 \mathcal{V}_{\alpha \beta \gamma}(\vec{q},\vec{k},\vec{p}) =
  \frac{(\vec{q}\vec{k})}{q}c_{\alpha \beta}(k) \delta_{\alpha \gamma}
  +\frac{(\vec{q}\vec{p})}{q}c_{\alpha \gamma}(p) \delta_{\alpha \beta}\,.
\label{vertex}
\end{equation}
Here we have additionally approximated static three-point correlation
functions in terms of two-point ones. Their inclusion is computationally
too demanding and does not change the results strongly in dense,
non-network forming systems such as ours \cite{sciortino2001}.
$\bm{c}(q)$ is the matrix of direct correlation functions defined through
the Ornstein-Zernike equation
\begin{equation}
  S_{\alpha \beta}^{-1}(q)
  =\delta_{\alpha \beta}/x_\alpha-\rho c_{\alpha \beta}(q)\,.
\end{equation}
Thus, taking $\bm{S}(q)$ from simulation, the collective dynamical density
correlators are fully determined in the theory.

The tagged-particle correlator $\Phi^s_\alpha(q,t)$ for a singled-out
particle of species $\alpha$ obeys an equation
similar to Eq.~\eqref{timespace_final},
\begin{multline}
 \frac{1}{\Omega^s(q)^2} \ddot{\Phi}_\alpha^s(q,t)+
  \frac{\nu^s(q)}{\Omega^s(q)^2}\dot{\Phi}_\alpha^s(q,t) +\Phi^s_\alpha(q,t)
\\
 +\int_{0}^{t} dt' \, M^s_\alpha(q,t-t')\Phi^s_\alpha (q,t') =0
\label{tagged}
\end{multline}
with $\Omega^s(q)^2=q^2 v_{\text{th},s}^2$
and the initial conditions $\Phi^s_\alpha(q,t=0)=1$, $\dot{\Phi}^s_\alpha(q,t=0)=0$. We set $\nu^s(q)=\nu$ to obtain damped-Newtonian short-time dynamics.
The corresponding memory kernel is evaluated from Eqs.~\eqref{memorykernel}
and \eqref{vertex} by considering a (M+1)-component mixture in the limit of
one concentration going to zero:
\begin{multline}
  M^s_\alpha(q,t) = \mathcal{F}^s_\alpha\left[ \bm{\Phi}(q,t),{\Phi}^s(q,t)\right]  = \\ \frac{\rho}{q^2} \sum_{\alpha'\beta'\neq s}
  \int \frac{d^3k}{(2 \pi)^3} \mathcal{V}_{s, \alpha'\beta'}(\vec{q},\vec{k})\,
  \Phi_{\alpha'\beta'}(k,t) \Phi^s_\alpha(p,t),
\label{vertex_tagged}
\end{multline}
with the tagged-particle vertex
\begin{equation}
\mathcal{V}_{s, \alpha \beta} (\vec{q},\vec{k})
  = \frac{\vec{q} \vec{k}}{q} c_{s \alpha}(k) c_{s \beta}(k).
\end{equation}
As the
memory kernel in Eq.~\eqref{vertex_tagged} contains the coherent correlator,
solving Eq.~\eqref{tagged} requires the full coherent dynamics to be known.

By virtue of the expansion
$\Phi^s_\alpha(q,t)=1-q^2 \delta r_{s,\alpha}^2(t)/6+\mathcal{O} (q^4)$,
the mean-squared displacement of species $\alpha$ is connected to the tagged-particle correlator via
\begin{equation}
\delta r_{s,\alpha}^2(t)=\lim_{q \to 0} \frac{6}{q^2}[1-\Phi^s_\alpha(q,t)]\,.
\end{equation}
Its equation of motion follows from Eq.~\eqref{tagged}:
\begin{equation}
 \partial_t \delta r^2_{s,\alpha}(t)  + v_{\text{th},\alpha}
 \int_0^t  dt' \, \hat{M}^s_\alpha (t-t') \delta r^2_{s,\alpha}(t')
  = 6 v_{\text{th}^2, \alpha} t \,,
\label{msd_eq}
\end{equation}
with
$\hat{M}^s_\alpha(t)= \lim_{q \to 0}  q^2 M^s_\alpha(q,t)$ and
\begin{multline}\label{msd_eq_m}
\hat{M}^s_\alpha(t)= \frac{1}{6 \pi^2} \sum_{\alpha'\beta'\neq s}
  \int_0^\infty dk \, k^4 c_{s \alpha'}(k)  c_{s \beta'} (k)
  \\ \times \Phi_{\alpha'\beta'}(k,t) \Phi^s_\alpha(k,t)
\end{multline}
Equation \eqref{msd_eq_m} states that the mean-squared displacement is
completely determined from the collective and tagged-particle density
correlation functions; there is no back-coupling of the MSD to itself
since the phase space at $q=0$ has vanishing contribution inside the
integral.

Many features of the solutions of the above MCT equations are known,
especially concerning points asymptotically close to MCT glass
transitions. We only summarize the basic results for completeness,
referring to the literature \cite{leshouchesI,goetzeBuch,Franosch,Fuchs1998}
for details. The starting point of the asymptotic analysis is to realize
that the MCT equations allow for bifurcation points for the long-time
limit of their solutions. Denoting them by
\begin{align}
  \bm {F}(q)&= \lim_{t \to \infty} \bm{\Phi}(q,t) &
  F^{s}(q) &= \lim_{t \to \infty} \Phi^s(q,t)\,,
\end{align}
one finds that these long-time limits (synonymously called glass form
factors or nonergodicity factors) are determined by a set of coupled,
implicit nonlinear equations,
\begin{gather}
  \label{finding_f}
  \bm{S}(q) - \bm{F}(q) =
  \left[\bm{S}^{-1}(q) + \mathcal{F}\left[{\bm F}(q) \right] \right]^{-1}\,,
\\
  \label{finding_fs}
  \frac{F^{s}(q) }{1-F^{s}(q)} = \mathcal{F}^s[{\bm F}(q), F^{s}(q)]\,.
\end{gather}
In the usual case, it is Eq.~\eqref{finding_f} that displays bifurcations:
Out of the possibly many solutions to this equation, the long-time
limit corresponds to the non-negative real solution that is largest
according to a straightforward ordering defined for each $q$ separately
and through the positive-definiteness relation
\cite{goetze_sjoe2,franosch2002}. The glass transitions of MCT are
then the bifurcation points affecting this largest solution
that arise from smooth changes in
the control parameters, and the most common case is that
of a $\mathcal{A}_2$ bifurcation where the long-time limit jumps
discontinuously from the trivial zero solution indicating a liquid
to a finite value indicating a solid.
The solution $F^s(q)$ determined as the largest solution of
Eq.~\eqref{finding_fs} will then, in the generic case, inherit the
bifurcations of ${\bm F}(q)$, and we will not discuss the possibility of
extra singularities arising in Eq.~\eqref{finding_fs} itself.
The generic case is in particular obeyed by the problem at hand,
quasi-hard-sphere tracer particles inside a quasi-hard-sphere system composed
of particles of roughly equal size.
Generally, the transition points are then defined as the points where the
stability matrix $\mathcal{C}$ of the nonlinear Eq.~\eqref{finding_f}, given by
\begin{eqnarray}\mathcal{C}[{\bm H}(q)] := 2(\bm{S}^c(q) - \bm{F}^c(q)) \times 
\nonumber
\\
\mathcal{F} \left[\bm{F}^c(q), \bm{H}(q) \right] ( \bm{S}^c(q) - \bm{F}^c(q))\,,
\label{stability_matrix}
\end{eqnarray}
has a unique critical eigenvector $\bm{H}(q)$ with eigenvalue unity.
$\bm{H}(q)$ is also called the critical amplitude (up to trivial
normalization factors that are sometimes split off from it).
We will denote quantities corresponding to such a transition point with
superscript, e.g., $\bm{F}_q^c$ and $F_q^{s,c}$.

On the liquid side of the transition, correlation functions follow a two step
relaxation scenario: for times large compared to the characteristic time
of the short-time motion, $t\gg t_0$, they decay with a time fractal
$\sim t^{-a}$ to the plateau, which extends until the $\beta$-relaxation time
scale $t_\sigma$. For $t\gg t_\sigma$, the decay from the plateau sets in
with the von~Schweidler law, $\sim-t^b$, initiating the final $\alpha$
relaxation that is characterized by a second time scale $t_\sigma'$.
The asymptotic analysis proceeds by analysing the equation of structural
relaxation, where time-derivatives that affect only the short-time motion
have been dropped,
\begin{eqnarray}
\bm{\Phi}(q,t)  &=& \bm{S}(q) \bm{M}(q,t) \bm{S}(q) 
\nonumber
\\
&-&\frac{d}{dt} \bm{S}(q)\int_{0}^{t} dt' \, \bm{M}(q,t-t') \bm{\Phi}(q,t')\,.
\label{notimederivatives}
\end{eqnarray}
Identifying the distance of the correlator to its plateau value as a small
parameter $\sigma$, one extracts the two time scales that diverge upon
letting $\sigma\to0$,
\begin{align}
 t_\sigma &= t_0 |\sigma|^{-1/2a}\,, &
 t_\sigma'=t_0 B^{1/b} |\sigma|^{-\gamma}\,,
\end{align}
where $\gamma=1/(2a)+1/(2b)$, and $a>0$ and $b>0$ are nontrivial and
nonuniversal exponents determined by the details of the interaction potential (see below).
The separation parameter $\sigma$ is, in leading order, linearly connected
to the change in control parameters, viz.\ in our case
$\sigma = C \epsilon+\mathcal{O}(\epsilon^2)$ with
$\epsilon=(\varphi-\varphi^c)/\varphi^c$. By convention, $\epsilon<0$
indicates a liquid state, $\epsilon>0$ the glass.
Note that $t_\sigma'/t_\sigma$ also diverges as $\sigma\to0$, so that
asymptotically close to the MCT transition, an ever larger window for
structural relaxation around the plateau open. In practice, this window
is cut short for large $-\epsilon$ by preasymptotic corrections, and for
small $|\epsilon|$ when the theory fails to describe residual ergodicity-restoring
processes in the glass.

For times $\hat{t}= t/t_\sigma$ where the correlator $\bm{\Phi}(q,t)$ is
close to $\bm{F}^c(q)$ one makes the following Ansatz
\begin{equation}
\bm{\Phi}(q,\hat{t}t_\sigma) = \bm{F}^c(q) + \sqrt{\sigma} \bm{G}(q, \hat{t}) + \mathcal{O}(|\sigma|)\,,
\label{betascaling}
\end{equation}
and the uniqueness of the critical eigenvector at the bifurcation point
implies the factorization theorem,
$\bm{G}(q, \hat{t}) = \bm{H}_q \cdot G(\hat{t})$. The function $G(\hat{t})$ is
determined by the so-called $\beta$-scaling equation
\begin{equation}
\frac{d}{d \hat{t}} \int_{0}^{\hat{t}} dt' G(\hat{t}-t')G(t') -\lambda (G(\hat{t}))^2 + {\sgn} \sigma =0\,.
\end{equation}
The nonuniversal details of the vertices enter in this equation only thorugh the
exponent parameter,
\begin{eqnarray}
\lambda &=&  \hat{\bm{H}}_q : \left( \bm{S}^c(q) - \bm{F}^c(q)\right)
  \mathcal{F}^c(q)\left[\bm{H}(q), \bm{H}(q) \right] \times
\nonumber
\\
 && \left( \bm{S}^c(q) - \bm{F}^c(q)\right) / \mathcal{N}
\end{eqnarray}
where $\mathcal{N}=\hat{\bm{H}}(q):(\bm{H}(q) ( \bm{S}^c(q) - \bm{F}^c(q))^{-1}\bm{H}(q))$,
and the double-dot operator includes contraction over $q$.
Here, $\hat{\bm{H}}(q)$ is the left-eigenvector corresponding
to $\bm{H}(q)$.

For times $t_0 \ll t \ll  t_\sigma$ the decay to the plateau at the critical
point is then governed by the $\beta$-relaxation; in leading order,
\begin{equation}
 \bm{\Phi}(q,t) = \bm{F}^c(q) + \bm{H}(q)(t/t_\sigma)^{-a}+ \mathcal{O}\left( (t/t_\sigma)^{-2a} \right)
\label{toplateau}
\end{equation}
where $a$ is determined as solution $0<x=a<1/2$ of
\begin{equation} \label{exponent}
\lambda = \frac{\Gamma^2 (1-x) }{\Gamma(1-2x)}.
\end{equation}
For times $t_\sigma \ll t \ll t_{\sigma'}$ and for $\sigma < 0$ the decay of
the correlator is described by the von~Schweidler law
\begin{equation}\label{vsmix}
 \bm{\Phi}(q,\hat t) = \bm{F}^c(q) - \bm{H}(q)(t/t_\sigma)^{b} + \mathcal{O}\left( (t/t_\sigma)^{2b} \right).
\end{equation}
Here $b$, the von Schweidler exponent is determined from the negative
solution $1>b=-x>0$ of Eq.~\eqref{exponent}.

In the glass ($\sigma >0$) the nonergodicity parameters behave like
\begin{equation}
 \bm{F}(q) = \bm{F}^c(q) + \bm{H}(q) \sqrt{\frac{\sigma}{1-\lambda}}
  +{\mathcal O}(\sigma)\,.
\end{equation}
Again we define the ``polydispersity-averaged'' total nonergodicity parameters
and the total critical amplitudes by summing over the bins,
\begin{equation}\label{summedfqhq}
f_q^c = \frac{ \sum_{\alpha \beta}  F^c_{\alpha \beta}(q)}
  {\sum_{\alpha \beta} S_{\alpha \beta}(q)},
  \; \; \; h_q = \frac{ \sum_{\alpha \beta}  H_{\alpha \beta}(q)}
  {\sum_{\alpha \beta} S_{\alpha \beta}(q)}.
\end{equation}
For timescales $ \tilde{t} =  t/  t'_\sigma \sim 1$ and $\varphi \to \varphi^c$ the $\alpha$-master equation can be derived
\begin{eqnarray} \tilde{ \bm{\Phi}}(q,\tilde t)  &=& \bm{S}(q) \bm{M}(q, \tilde t) \bm{S}(q )
\nonumber
\\
&-& \frac{d}{d \tilde t} \bm{S}(q) \int_{0}^{\tilde t} dt' \,
  \bm{M}(q,\tilde t-t') \tilde{\bm{\Phi}}(q, t').
\label{alphamaster}
\end{eqnarray}
This equation states that all correlators should collapse on the same function
when rescaled by an appropriate scaling time $\tilde t$. This is due to the
invariance of Eq.~\eqref{alphamaster} when rescaled in time and is the
mathematical manifestation of the time-temperature superposition principle.

\subsection{Numerical details of the MCT solution}

MCT calculations were performed with static structure factor matrices taken
from the simulations; to access packing fractions for which no simulations
have been run, linear interpolation in $\varphi$ was used for $\bm{S}(q)$.
As in the simulation, calculations with a one-component ($M=1$), and two
different multi-component binnings, $M=3$ and $M=5$, have been performed
(see Sec.~\ref{antoniotext}). To choose effective diameters $d_\alpha$ for
the different bins, we have followed the moment approximation
\cite{goetze1987b,voigtmann2003}: for $M=3$, the three values $\{d_1,d_2,d_3\}$ and equal concentrations allow
to match the actual polydispersity distribution up to the second moment
(with the requirement that the skewness of the distribution vanishes),
and for $M=5$, additionally the fourth central moment can be matched using equal concentrations.
This leads to the choice $\{d_1,d_2,d_3\}=\{1-1/\sqrt{200},1,1+1/\sqrt{200}\}$
in the $M=3$ system, and $\{d_1,d_2,d_3,d_4,d_5\}
=\{0.91675, 0.96254, 1, 1.03745, 1.08325 \}$ for the $M=5$ system.

The problem of solving Eq.~\eqref{timespace_final} in full has to be tackled
numerically, choosing a suitable discretization. The wave number grid
selected here is $q \in [0.1, 60.0]$ in steps of $\Delta q = 0.2$. The
discretization is a compromise between calculation time and being as close as
possible to the simulation structure factors. The initial time step was chosen
as $\delta t=10^{-6}$. After every $128$ steps the stepsize was doubled in
order to cover logarithmically large intervals in $t$.

\section{Data Analysis}\label{asymptotic}

\subsection{Structure factors}

\begin{figure}
\includegraphics[width=.9\linewidth]{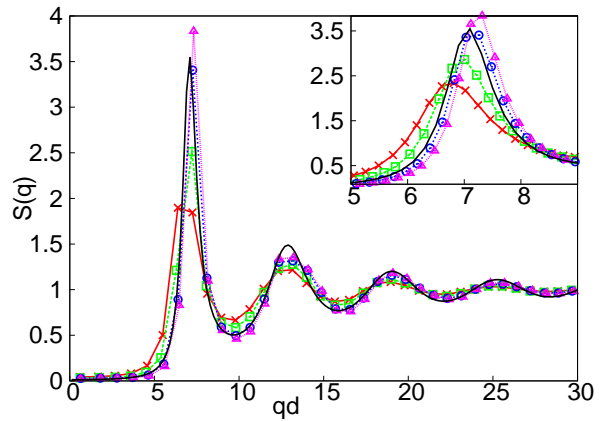}
\caption{\coloronline
  Total static structure factor extracted from the simulation.
  The packing fractions $\varphi=0.45$, $0.5$, $0.55$, and
  $0.57$ are marked with crosses (red), squares (green), circles (blue) and
  triangles (magenta), respectively. The black solid line is the Percus-Yevick
  result for $\varphi = 0.516$. The inset shows a magnification of the peak.}
\label{sq1cp}
\end{figure}

We first address the static structure factor that serves as an
input to MCT, ${\bm S}(q)$ and $S(q) = \sum _{\alpha \beta} S_{\alpha \beta} (q)$ 
obtained from the simulations. In Fig.~\ref{sq1cp}, the averaged structure
factor $S(q)$ is shown for different packing fractions. It exhibits the
standard features known for dense liquids where excluded volume is the main
interaction effect: a pronounced first peak is visible that shifts to
higher $q$-values and increases in intensity with increasing packing
fraction, reflecting a decreasing average next-neighbor distance and increased
local ordering in the denser system. The figure also includes $S(q)$ as calculated
from the Percus-Yevick (PY) approximation for monodisperse hard spheres
\cite{wertheim1963}. It is known that this approximation performs quite well,
but somewhat over-estimates the amount of ordering present in the system;
this is expressed by a shift in density values. Still, comparing densities
where the first main peak is reasonably well described within PY, the
amplitude of the second peak in $S(q)$ is notably overestimated by the
approximation in comparison to our simulation results.
Note that in MCT, the main contribution to the memory kernel causing slowing
down and arrest comes from these amplitudes in $S(q)-1$. As larger
wave numbers contribute more strongly to the three-dimensional integral,
not just the main peak of $S(q)$, but also the shape of its large-$q$
envelope determine the MCT dynamics; one can thus anticipate from
Fig.~\ref{sq1cp} that MCT calculations based on the PY structure factor
will clearly underestimate the critical packing fraction.

\begin{figure}
\includegraphics[width=.9\linewidth]{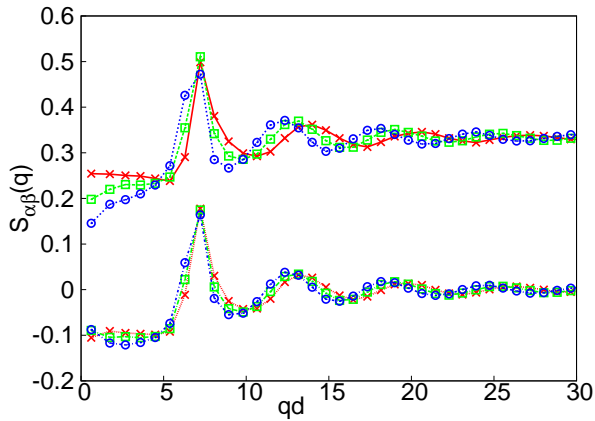}
\caption{\coloronline
  Partial static structure factors extracted from the simulation by
  binning the polydisperse system into $M=3$ species, at packing fraction
  $\varphi = 0.5$. The upper curves show the diagonal terms
  $S_{11}(q)$ (red crosses), $S_{22}(q)$ (green squares), $S_{33}(q)$
  (blue circles). The lower curves  show the off-diagonal terms $S_{12}(q)$
  (red crosses), $S_{13}(q)$ (green squares) and $S_{23}(q)$ (blue circles).}
\label{sq3cp}
\end{figure}

In Fig.~\ref{sq3cp}, we show the partial structure factors resulting from
a binning of the simulation data into a
three-component system. All diagonal terms as well as the off-diagonal
terms (except for a trivial shift by the constant 
$1/(x_\alpha x_\beta)^{1/2}=1/3$) are
very similar as expected for a mixture of almost equal constituents.
The slightly different average next-neighbor distances
for the different particle sizes cause corresponding shifts in the oscillation
frequencies and hence the positions of the peaks in $S_{\alpha\beta}(q)$.
This effect is most pronounced in the diagonal terms, as these only contain
information from one distinct particle-size bin.
Adding these partial structure factors recovers the $S(q)$ shown
in Fig.~\ref{sq1cp}, and elucidates that the comparatively weak
high-$q$ peaks visible there are the result of a destructive interference
of the slightly shifted oscillations in the multi-component $\bm{S}(q)$.
However, the main peak is strong enough to be less affected.
Thus, for systems whose interactions are close to hard-core repulsion, such as
our system, averaging $S(q)$  mainly results in an underestimation of the coupling
strength, i.e., too fast dynamics in the one-component calculation as
compared to a multi-component calculation. Indeed, our MCT calculations
discussed below essentially confirm this expectation, that in the range
of $q$ around the first peak in $S(q)$ and above, the changes between
pre- and post-averaging are minor once the shift of the critical packing
fraction is acocunted for. Yet, for small $q$, the situation is more
intricate. Let us also note that for systems whose glass transition
features are dominated by the large-$q$ behavior of the static structure factor,
the destructive interference induced by pre-averaging the MCT input can
lead to severe changes in the qualitative dynamics \cite{henrich2007}.
This particularly concerns polydisperse colloid-polymer systems with
short-ranged (depletion-induced) interaction among the colloids.

\subsection{Critical point density}

\begin{table}
\begin{center}
\begin{tabular}{p{.15\linewidth}
p{.15\linewidth}
p{.15\linewidth}
p{.15\linewidth}
p{.15\linewidth}}
model & $\varphi^c$ &  $\lambda$  & $b(\lambda)$ & $a(\lambda)$\\ 
\hline
$M=1$ & 0.566 &  0.717 & 0.613 & 0.320 \\ 
$M=3$ & 0.537  & 0.735 & 0.583 & 0.315\\ 
$M=5$ & 0.535  & 0.739 & 0.576 & 0.310\\ 
\end{tabular}
\caption{Table of MCT-calculated critical packing fractions for the
  different polydispersity models using $M$ bins. Calculations are based
  on the computer-simulated partial static structure factors.}
  \label{tabstat}
\end{center}
\end{table}

The critical packing fraction $\varphi^c$ of the MCT glass transition was
calculated with a bisection algorithm which returns the point where the
eigenvalue of the stability matrix $\mathcal{C}$, Eq.~\eqref{stability_matrix},
is unity. Table~\ref{tabstat} summarizes the values of $\varphi^c$ along
with the corresponding exponent parameters $\lambda$ and the MCT exponents
corresponding to these values, for the $M=1$, $3$, and $5$ calculations
we performed.

As expected from the discussion of the static structure factor above,
the glass transition point in the calculations shifts to lower densities
with increasing $M$. Remarkably, the most pronounced change already
occurs between $M=1$ and $M=3$, whereas the further increase to $M=5$
does not alter $\varphi^c$ or $\lambda$ substantially. We thus consider
$M=5$ to be already close to the limit of many components that should
in fact be taken to describe a truly polydisperse system. At this point we
would like to stress that the downshift of the glass transition point for
$M=1,3,5$ is only related to the different MCT-models used here, as the 
polydispersity in the simulation is not changed.
It should be noted
that still, even the value $\varphi^c_{M=5}\approx0.535$ is significantly
above the value calculated within the PY approximation,
$\varphi^c_\text{PY}\approx0.516$ \cite{Franosch}. However, the MCT glass
transition from an asymptotic analysis of the simulation data yields
a value $\varphi^c_\text{MD}\approx0.59$ \cite{voigtmann2004},
much higher than the values
calculated within the theory even with the correct static structure
information, but in reasonable agreement with the commonly quoted value
for polydisperse hard-sphere like colloidal suspensions,
$\varphi^c_\text{exp}\approx0.58$ \cite{vanMegen}.
This is a well known shortcoming of MCT, and we consider the
better agreement we obtain for the one-component calculation,
$\varphi^c_{M=1}\approx0.566$ as fortuitous. Note also that one usually
expects polydispersity to shift the glass transition to higher densities,
as the overall packing efficiency incresases \cite{braun}. However,
the role played by the different shapes of the polydispersity distributions
is not well studied. Within MCT, this commonly observed trend of increasing
$\varphi^c$ in mixtures is, at least for binary hard-sphere mixtures,
only predicted for size ratios $\delta\lesssim0.7$, while for $\delta$ closer
to unity, the inverse trend is found \cite{voigtmann2003}. Relating
$\delta$ to the extreme $d_\alpha$ in our calculations, even our $M=5$
system only corresponds to $\delta\approx0.85$.

\subsection{$\alpha$-process analysis} \label{alpha-section}

Next, we test the validity of $\alpha$ scaling for our simulation data.
According to Eq.~\eqref{alphamaster}, plotting correlators as functions
of $t/t_\sigma'$ should collapse the data for long times, with a master
curve extending as $\varphi\to\varphi^c$ from below.

\begin{figure}
\includegraphics[width=.9\linewidth]{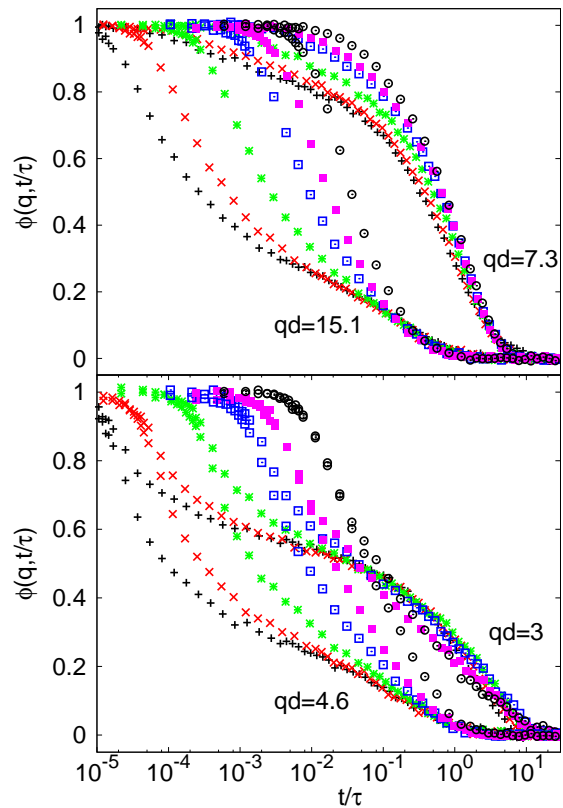}
\caption{\coloronline
  Simulation correlators rescaled by the $\alpha$ timescale $\tau
  \propto t'_\sigma$ for four different wave numbers $q$ as indicated,
  where $\tau$ was determined by shifting the results for lower $\varphi$
  to agree at long times with the $\varphi=0.585$ curve at $qd=7$, and
  does not depend on $q$.
  Packing fractions shown are $\varphi = 0.585$ (black plus symbols),
  $\varphi = 0.58$ (red crosses), $\varphi = 0.57$ (green stars),
  $\varphi = 0.55$ (blue open squares), $\varphi = 0.53$
  (magenta filled squares) and $\varphi = 0.50$ (black circles).}
\label{rescaled-cor}
\end{figure}

To determine a relaxation time $\tau\propto t_\sigma'$ from the data alone,
we have shifted the correlation functions at a
single fixed value of $q=7.3/d$ to coincide with the corresponding
curve
at the highest packing fraction in the liquid, $\varphi=0.585$,
at long times. After this procedure,
the validity of the scaling can be checked by requiring that $\tau$
be independent on $q$. Note that $qd\approx7.3$ corresponds to the position
of the main
peak in the averaged static structure factor. We have chosen this value
since here the
strength of the $\alpha$ process is maximal, and the best separation from
the $\beta$ regime is achieved. Fig.~\ref{rescaled-cor} shows the result
of this scaling for four different wave numbers.
An $\alpha$-master curve clearly is approached,
with the scaling regime for the highest two densities extending over
about two orders of magnitude in time. The strong coupling of the $\alpha$-relaxation on local length scales predicted by MCT is observed, as the same scaling factor $\tau$ rescales the correlators for different wavevectors.

In the lower panel of Fig.~\ref{rescaled-cor}, the $\alpha$ scaling is
exhibited for small $qd$; here, deviations are visible at $qd=3$, where the
simulation data does not exhibit a common shape for the $\alpha$-relaxation
part of the correlators. The highest $\varphi$ shown indicate
a decay that is either more stretched or exhibits a further inflection point
in the $\phi(\tilde t)$-versus-$\log\tilde t$ plot below the plateau.
We will return to such features below in the discussion of hydrodynamic
interdiffusion modes that exist in multi-component systems and interplay with
the structural relaxation at low $qd$.

A common description of the shape of the $\alpha$ relaxation is in terms of
stretched-exponential (Kohlrausch) laws,
\begin{equation}
 \Phi(q,t) \approx A_q \exp[-(t/\tau_q)^{\beta_{q}}]\,.
\end{equation}
Here, $\beta_q$ is the stretching index, required to be $\beta_q\le1$ for
structural relaxation in equilibrium systems. While the $\alpha$-master
function from MCT is different from the Kohlrausch form, the theory predicts
that for large wave numbers, the two functional forms become identical,
and $\beta_{q\to\infty}\to b$ \cite{fuchs_non_cryst}.
$\tau_q$ is commonly referred
to as the $\alpha$-relaxation time, and it is connected by a $q$-dependent but density-independent 
prefactor to the scaling time $t_\sigma'$ appearing in MCT.
The Kohlrausch amplitude $A_q<1$ can be taken as an estimate of the
MCT plateau value $f^c_q$, and since the $\alpha$ process by definition starts
below this plateau, $A_q\le f^c_q$ is expected. In practice, however,
the separation of the $\alpha$ process from the $\beta$ relaxation is
often not clear enough to warrant this restriction.

\begin{figure}
\includegraphics[width=.9\linewidth]{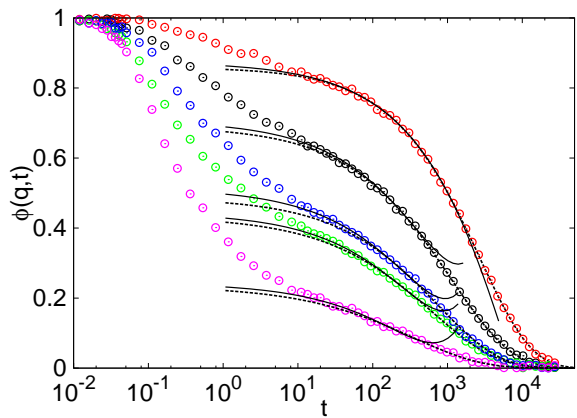}
\caption{\coloronline
  Fits of stretched-exponential Kohlrausch functions (dashed black lines)
  to the simulated coherent density correlators (circles) at
  $\varphi =0.585$. The q-values are from top to bottom $qd= 6.6$ (red),
  $7.4$ (black), $9.8$ (blue), $12.8$ (green) and $15.6$ (magenta). The fit
  range was chosen as $t\in[100:10^5]$. Solid black lines show von~Schweidler
  fits up to second order, Eq.~\eqref{vsfits},
  with $t_\sigma =1000$ and $b=0.53$, fitted in the range $t\in[20:336]$.}
\label{kohl}
\end{figure}

In general, Kohlrausch fits are hindered by some subtle problems that are
often overlooked. Lacking a clear separation of the $\alpha$ process,
the fit parameters exhibit a dependence on the fit range. It is not clear
a~priori how to choose the optimal fit range, as for very long times, one
expects the relaxation to become (non-stretched) exponential again (and to
be covered within the noise of any experiment),
and for short times, deviations are seen that can be understood within MCT
to be due to the difference between $\beta_q$ and $b$ for finite $q$.

We have tried to fix the fit range of our stretched-exponential fits such that
the fit parameters exhibit only a weak dependence on the fit boundaries.
Examining the $q=q_p$ correlators for $\varphi=0.585$, this leads to
$t\in[10^2,10^5]$. In Fig.~\ref{kohl}, examples of such fits are shown.
Although the
agreement is generally convincing, let us stress that these fits are a
pure data analysis, not taking into account the full numeric MCT calculations
we will discuss below. The curves in Fig.~\ref{kohl} show what one can
extract parameters like the plateau value or the von Schweidler exponent $b$ \emph{without} having performed full MCT calculations.
In Fig.~\ref{kohl}, we additionally show the results of von~Schweidler
fits to the simulation data: adapting Eq.~\eqref{vsmix} and extending
it by the next-to-leading order \cite{Franosch}, we fitted
\begin{equation}\label{vsfits}
\Phi(q,t)=f_q-h_q (t/t_\sigma)^b\cdot \left(1-k_q(t/t_\sigma)^b \right)\,,
\end{equation}
determining $f_q$, $h_q$, and $k_q$ by fitting and fixing $b=0.53$ to be
consistent with the analysis of $\alpha$-relaxation stretching presented
below (Fig.~\ref{betaall}).
Note that a free fit of von~Schweidler's law without fixing $b$ is usually
ambiguous as the determination of the MCT exponents directly from data
bears uncertainties \cite{sciortino1999}.
Eq.~\eqref{vsfits} has been used in the range $t\in[20; 336]$. The
von~Schweidler results then represent about two decades in time of the
correlation functions.

\begin{figure}
\includegraphics[width=.9\linewidth]{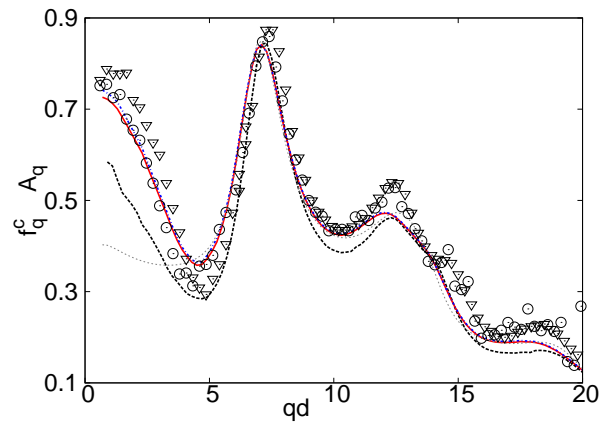}
\caption{\coloronline
  MCT results for the critical nonergodicity parameters $f^c_q$ obtained by
  binning the simulated static structure-factor data into $M=1$ (black dashed),
  $M=3$ (red solid) and $M=5$ components (blue dash-dotted). The grey
  dotted curve is the MCT solution with the Percus Yevick static
  structure factor ($M=1$). Circles show the plateau values of the simulation
  curves obtained by fitting Kohlrausch functions to the coherent simulation
  correlators at $\varphi=0.585$ (Fig.~\ref{kohl}). Triangles are obtained by
  von~Schweidler fits (Eq.~\eqref{vsfits} and Sec.~\ref{beta-process}).
}
\label{fqall}
\end{figure}

Figure~\ref{fqall} shows the results for the amplitude $A_q$, together with
the nonergodicity parameters $f^c_q$ calculated from MCT, and the estimates
of $f^c_q$ obtained from von~Schweidler fits. The values obtained from all
three methods in general show good agreement, although some details warrant
discussion. Let us first turn to the large-$q$ regime, $qd\gtrsim7$.
Here, the von~Schweidler fits, Eq.~\eqref{vsfits},
yield $f^c_q$ that are in good agreement
with the Kohlrausch amplitudes $A_q$, and the relation $A_q\le f^c_q$ is
reasonably well fulfilled. The MCT calculations somewhat underestimate
$f^c_q$, although the agreement is soupcon improved for the multi-component
systems with $M=3$ and $M=5$, in particular for values of $q$ where the
averaged static structure factor (and hence also the $f^c_q$-versus-$q$
curve) shows minima. The underestimation of $f^c_q$ by MCT can be
attributed to the underestimation of $\varphi^c$: MCT describes
arrest at lower densities, but $f_q$ increases with density as the denser
glass is stiffer with respect to density fluctuations.
For comparison, we also show in Fig.~\ref{fqall} the $f^c_q$ obtained by
employing the PY approximation within MCT; this result agrees well
with the $M=3$ and $M=5$ calculations,
indicating in particular that for the arrest of
small-wavelength fluctuations, polydispersity and particle softness do not
play a major role. One in fact recognizes that the truly monodisperse
calculation (with PY input) is in better agreement with the
polydisperse data than the one-component calculation using the
polydispersity-averaged $S(q)$. This can be intuitively interpreted:
pre-averaging in $S(q)$ artifically reduces large-$q$ static correlations,
but $f_q$ is in essence a dynamical quantity, and the effect
of the dynamics on these large-$q$ correlations needs to be included to
describe the degree of dynamical arrest. Still, in Fig.~\ref{fqall} this
is a merely quantitative effect.

For the large-wavelength regime, $qd\lesssim6$, the situation is more
differentiated: here, both one-component calculations differ more notably
from the results obtained by pure data fitting, and the results calculated
with the $M=3$ or $M=5$ theory. In particular, the PY-based one-component
result shows a rather weak $q$-dependence of the plateau value for small $q$,
and yields $f^c_q\approx0.4$ in this regime. In contrast, the multi-component
results show an upturn of $f^c_q$ as $q$ is decreased, with plateau values
rising to almost $0.8$ at the lowest $q$ accessible in the simulation.
The one-component MCT calculation based on the simulated (pre-averaged)
structure factor also shows this upturn, but less pronounced.

This behavior can be rationalized by the effect of polydispersity: first,
going over from the strictly one-component PY approximation to the simulated
$S(q)$ already contains static, pre-averaged information about the
size distribution, and this is sufficient to explain qualitatively that
polydispersity
tends to increase the stiffness of the frozen structure towards long-range
density fluctuations. The post-averaged calculation is needed to
capture this trend quantitatively. It shows that additionally the freezing out of the  interdiffusion process increases the average $f_q$ at small $q$ \cite{fuchs1993}. 

\begin{figure}
\includegraphics[width=.9\linewidth]{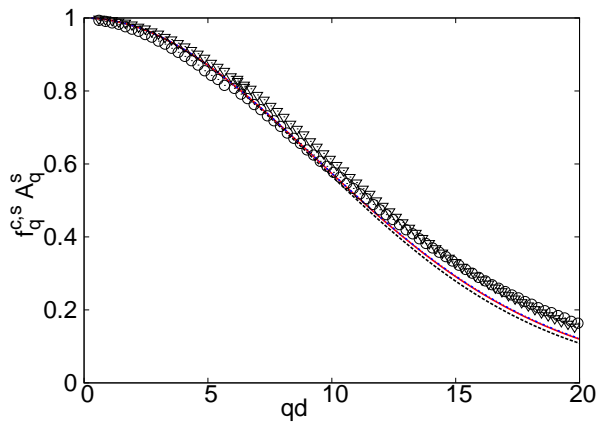}
\caption{\coloronline
  MCT results for the tagged-particle critical nonergodicity parameters
  $f^{s,c}_q$ for the $M=1$ (dashed), $M=3$, (solid) and $M=5$ (dash dotted)
  polydispersity models. The solutions for $M=3$ and $M=5$ cannot be
  distinguished on the scale fo the figure.
  Open circles and triangles show plateau values determined from fitting
  Kohlrausch and von~Schweidler functions, respectively, as in
  Fig.~\ref{fqall}.
\label{fq_inc}
}
\end{figure}

The tagged-particle analog of the quantities shown in Fig.~\ref{fqall},
$f^{s,c}_q$, is exhibited in Fig.~\ref{fq_inc}. Here, the differences
in the different methods of determining the plateau height are much smaller,
as intuitively expected; the incoherent $f^{s,c}(q)$-versus-$q$ shapes do
not show the oscillations typical for the collective quantities. Only at the
largest wave numbers investigated here, $qd\gtrsim12$, some differences
between the MCT calculation and the fit results become apparent. Here,
the theory again underestimates the amount of arrested density fluctuations,
and this can agian be rationalized by its lower critical density. For
small $q$, all $f^{s,c}(q)$ have to approach unity, so that any differences
are trivially wiped out.

\begin{figure}
\includegraphics[width=.9\linewidth]{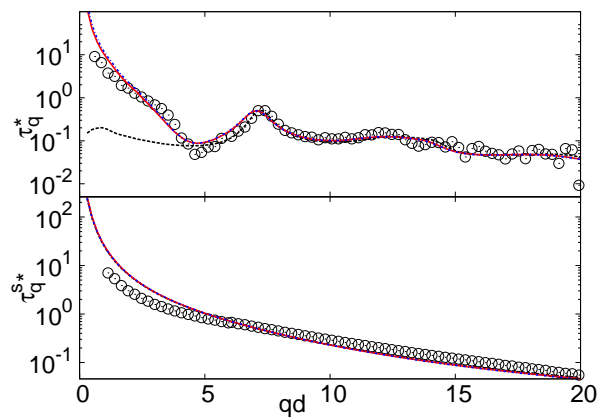}
\caption{\coloronline
  $\alpha$-relaxation times $\tau_q$ obtained by stretched-exponential fits to
  the MCT master curves for the $M=1$ (black dashed), $M=3$ (red solid), and
  $M=5$ (blue dash-dotted) polydispersity approximations. All values have
  been rescaled by their value at $q_pd=7.3$, $\tau^*=\tau_q/\tau_{q_p}$.
  The upper panel shows data corresponding to the collective density
  correlation functions, the lower panel those corresponding to the
  tagged-particle analog.
  Circles correspond to the $\tau_q$ extracted from the simulation data
  at $\varphi=0.585$, as in Fig.~\ref{fqall}.
}
\label{tauall}
\end{figure}

Let us now turn to a discussion of the $q$-dependence of the
relaxation time $\tau_q$ and the stretching $\beta_q$ resulting
from Kohlrausch fits. To obtain equivalent values also for the MCT curves,
we have fitted the theory's $\alpha$-master curves with stretched
exponential functions. Numerically, the master curves have been
approximated by correlators close to the transition point: choosing
$\epsilon=-10^{-6}$ in Eq.~\eqref{timespace_final} proves sufficient to
effectively solve Eq.~\eqref{notimederivatives} for the timescales of
interest.
For the simulation, we again restrict the discussion to the highest
density available, $\varphi=0.585$, which however corresponds to a
different $\epsilon$. Thus, to enable a meaningful comparison,
relaxation times are scaled to coincide at $q_pd=7.3$.

Figure~\ref{tauall} shows the resulting $\alpha$-relaxation times for both
the coherent (upper panel) and incoherent (lower panel) dynamics. A similar
distinction into two regimes as above arises: for $qd\gtrsim6$, the
$q$-dependence of the relaxation time is excellently predicted by the
theory. For $qd\lesssim6$, this holds for the collective correlation
functions only when comparing with the multi-component calculation. Even
the pre-averaged one-component calculation is in qualitative difference,
since there, $\tau_{q\to0}$ approaches a constant, resulting in a weak
$q$-dependence for all $qd\lesssim6$. Instead, in the multi-component
theory, $\tau_q\sim1/q^2$ as $q\to0$ for every single matrix element of
the partial-density correlation-function matrix. This has a clear physical
interpretation: $\tau_q\sim\text{const.}$ reflects the fact that the
overall momentum of the system is conserved. However, this is not the case
if one considers a single species inside a mixture of components only,
since momentum can be exchanged between the species. Only the sum of
the partial momenta is thus conserved, which is reflected in the existence
of an appropriate zero eigenvector in the MCT memory kernel; however, this
does not transcede to the averaged correlator defined by
Eq.~\eqref{summedphi} itself.
Strikingly, the resulting rise in $\tau_q$ as $q\to0$ that is apparent
in the $M\ge3$ MCT calculations is in very good agreement with the
simulation results, with only the lowest $q$ values of the simulation
deviating slightly (which may be due to minute finite size effects).

For the same reason (momentum conservation or rather the lack thereof
for a single species), $\tau^s_q\sim1/q^2$ is expected in any system.
Indeed, this is seen in all curves in the lower panel of Fig.~\ref{tauall}.
However, the divergence predicted by MCT is stronger than the one
seen in the simulation, an effect visible even at finite $q$. Only in the
large-$q$ regime defined above is the MCT description of the $\tau^s_q$
quantitatively accuracte. At the largest $q$ extracted from the simulation,
the theory in turn somewhat underestimates the relaxation times.

\begin{figure}
\includegraphics[width=.9\linewidth]{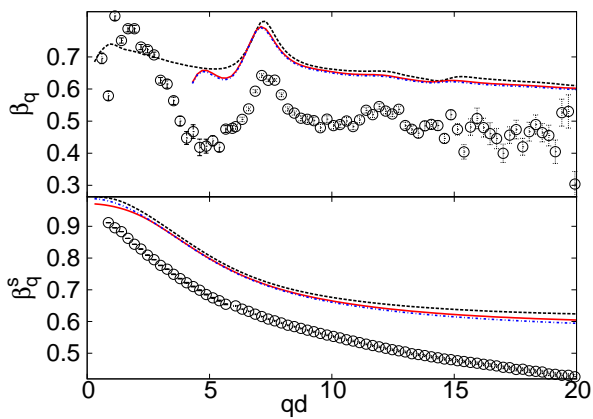}
\caption{\coloronline
  Kohlrausch stretching parameters $\beta_q$ as functions of wave number $q$,
  determined from fits to the MCT $\alpha$-master curve
  for the $M=1$ (dashed), $M=3$ (solid), and $M=5$ (dash-dotted) polydispersity
  moment approximations. $\beta_q$ determined from fits to the simulation
  data (as in Fig.~\ref{fqall}), are shown as open circles. The upper
  (lower) panel shows results from analyzing the collective (tagged-particle)
  quantities.
}
\label{betaall}
\end{figure}

Turning to the stretching exponents $\beta_q$, Fig.~\ref{betaall},
the agreement between MCT and simulation is less favorable. The values
obtained from fitting the theory curves are systematically too high,
an effect that will become evident also below when discussing the full
correlators. Only the qualitative behavior of $\beta_q$ with $q$ is
qualitatively in agreement, although the difficulty of determining
$\beta_q$ in the simulation for large $q$, where the amplitudes $A_q$ are
already rather low, does not allow a detailed discussion.

Since the $\alpha$ master curve strictly is a Kohlrausch function
only in the limit $q\to\infty$ \cite{fuchs_non_cryst}, emphasis should
mainly be placed on the behavior of $\beta_q$ at large $q$, and at small
$q$, as we will discuss below. For large $q$, the fits to the MCT curves
nicely exhibit the asymptotic behavior, $\beta_{q\to\infty}\to b$, with
a value of $b\approx0.6$, consistent with the values given in
Table~\ref{tabstat}. Also the simulation-fitted stretching exponents are
compatible with the approach to
a finite constant at large $q$, albeit with a somewhat lower value,
$b\approx0.5$; we can take this difference as an indication for the
error inherent in the value of $\lambda$ as calculated within MCT,
in particular since the value of $b$ is in good agreement with the
$b\approx0.53$ that results from an independent $\beta$-relaxation
analysis of the simulation data (see below).
For the simulated incoherent correlation functions, however, no such
large-$q$ limit can be identified for $\beta^s_q$; the reason for this
behavior is unclear. 

At low $q$, the appearance of a diffusion mode in the incoherent
correlator demands $\beta^s_q\to1$ for $q\to0$
(since hydrodynamic relaxation functions are just exponentials).
The fits to both the different theory calculations and to the simulation
data confirm this. For the coherent $\beta_q$, no such statement holds
in the true monodisperse system; there is no collective single-component
diffusion mode.
Hence, $\beta_{q\to0}<1$ is found for the MCT calculation based on the
pre-averaged structure factor. For the same reason discussed in connection
with the $\tau_q\sim1/q^2$ behavior above, however, the $\beta_q$ from fits
to the simulation data do exhibit an increase as $q$ decreases towards zero.
In principle, a similar trend should be found in the multi-component
MCT calculations. However, a peculiarity arises here, that prevents us
from determining meaningful values of $\beta_q$ in this case for
$qd\lesssim4$. Here, the appearance of a number of distinct interdiffusion
processes typical for a multicomponent system leads to $\alpha$-relaxation
curves that are superpositions of a small number of exponentials, leading
to master curves that exhibit multiple ``shoulders''; for the corresponding
relaxation spectra, this corresponds to multiple $\alpha$ peaks
\cite{diplomarbeit,fuchs1993}. Such interdiffusion processes in
principle also exist in the simulation; however in this truly polydisperse
system, a large number of them is combined to a relaxation curve that is
again reminiscent of a single $\alpha$ process (akin to a heterogeneous
superposition of single relaxators).

\subsection{$\beta$-process analysis} \label{beta-process}

\begin{figure}
\includegraphics[width=.9\linewidth]{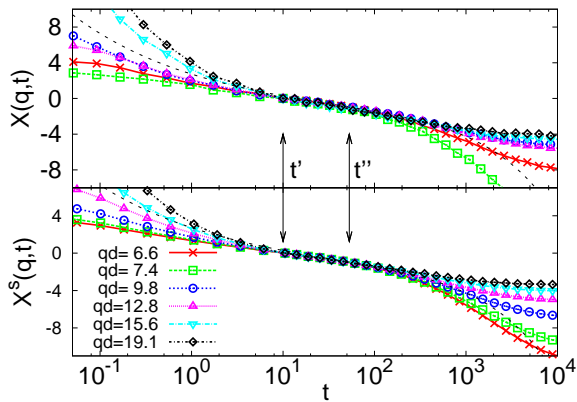}
\caption{\coloronline
  $\beta$ analysis of the simulation data at $\varphi = 0.585$. Functions
  $X(q,t)$ calculated from Eq.~\eqref{xeq} by fixing $t'=10.0085$ and
  $t''=52.5810$ are shown for the collective (tagged-particle) correlators
  in the upper (lower) panel. Different wave numbers $q$ were chosen as
  labeled. The MCT factorization theorem is validated by observing data
  collapse for different $q$ in a time window spanning $[t',t'']$, and by
  the $\beta$-master curve shown as a dashed line.
\label{xfunctions}
}
\end{figure}

We complete the asymptotic analysis of our simulation data by investigating the
$\beta$-scaling regime. On approaching $\varphi^c$, MCT states that the
correlation functions be described in leading order by Eq.~\eqref{betascaling}.
However, testing this relation involves a number of fit parameters whose
determination is difficult. An approach to testing the first main prediction
of MCT for the $\beta$-relaxation window, viz.\ the factorization theorem,
is to consider the function \cite{gleim2000}
\begin{equation}\label{xeq}
 X(q,t)= \frac{\phi(q,t)- \phi(q,t')}{\phi(q,t')-\phi(q,t'')}\,,
\end{equation}
with times $t'$ and $t''$ fixed to be inside the scaling regime.
Equation~\eqref{betascaling} then predicts $X(q,t)=x_1G(t)-x_2$ to be
independent on wave number; hence, superimposing the functions $X(q,t)$
for different $q$, one should be able to fix the two times $t'$ and $t''$
uniquely such that a time window appears where all $X(q,t)$ collapse.
The procedure has the advantage that the critical amplitude drops out and
thus does not need to be determined by fitting. It was shown to work
very reliably in a binary Lennard-Jones mixture by Gleim and Kob \cite{gleim2000}.

As shown in Fig.~\ref{xfunctions}, such a collapse is indeed possible
for the simulation data at $\varphi=0.585$, where we have chosen
$t'=10.0085$ and $t''=52.5810$. Both the collective and the tagged-particle
density correlators collapse for all the wavenumbers investigated for
a region spanning $[t',t'']$ and slightly extending to both smaller and
larger times.
Estimating that $\epsilon=-0.015$ ($-0.017$, $-0.020$) for the $M=1$ ($M=3$, $M=5$) 
analysis, one cannot expect the
first-order asymptotic result for the $\beta$-relaxation function to hold
over more than one decade in time \cite{Franosch}; indeed this is roughly
what we observe.

A stronger test of the MCT asymptotics implicit in Fig.~\ref{xfunctions} is
the so-called ordering rule: since in the next-to-leading order corrections
to the factorization theorem the same $q$-dependent correction
amplitudes appear both for the
early-time deviations and for the long-time corrections \cite{Franosch},
correlators that lie, say, above the $\beta$ correlator for short times
must also deviate in that direction for long times. Thus, numbering the
correlators in the order in which they deviate from the asymptote for
short times, the same numbering should be found on the long-time side.
Figure~\ref{xfunctions} confirms this. As was also found in
Ref.~\cite{Franosch} 
for the MCT calculations based on the
PY structure factor for hard spheres, this ordering rule is even preserved
among the correlators at long times, when the $\beta$ correlator already
violates it. This effect can be seen in Fig.~\ref{xfunctions} by noting
that the leading-order asymptote (drawn as a dashed line) emerges between
the $qd=12.8$ and $qd=15.6$ correlators, but already crosses the curves
for smaller $qd$ at around $t\approx500$, while the correlators obey the
ordering rule up to the time window plotted, $t\lesssim10^4$.

In order to extract the critical amplitude $h_q$ from the simulation,
one can define a function in analogy to Eq.~\eqref{xeq} by
\begin{equation}\label{yeq}
Y_{q} = \frac{\phi(q,t_1)- \phi(q, t_2)}{\phi(q_0,t_1)- \phi(q_0,t_2)} = \frac{h_q}{h_{q_0}}
\end{equation} 
with $t_1$, $t_2$ chosen in the $\beta$-scaling regime. The last equality
follows again from Eq.~\eqref{betascaling} and thus allows us to extract the
critical amplitudes up to a factor $h_{q_0}$. Since Eq.~\eqref{yeq}
becomes independent on the times chosen as long as they are in the
$\beta$-relaxation window, we can further improve the statistics of $Y_q$
by averaging over two time windows \cite{graeser2006}
\begin{equation}\label{Yfunc}
Y_{q} = \frac{ \sum_{j=1}^{n/2} \phi(q,t_j)-\sum_{j=n/2+1}^{n} \phi(q, t_j)}
  {\sum_{j=1}^{n/2} \phi(q_0,t_j)-\sum_{j=n/2+1}^{n} \phi(q_0,t_j)} =
  \frac{h_q}{h_{q_0}}
\end{equation} 
and use all the data points $t_j$ within the $\beta$-scaling regime, which
leads in our case for $\varphi=0.585$ to $t_j \in [10.0085; 52.5810]$.

\begin{figure}
\includegraphics[width=.9\linewidth]{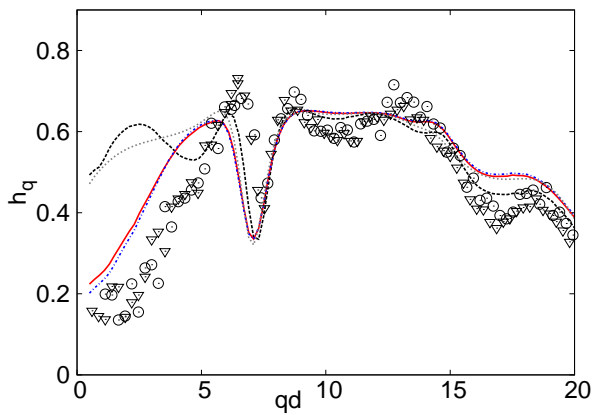}
\caption{\coloronline
  Critical amplitudes $h_q$ calculated within MCT for the $M=1$ (dashed),
  $M=3$ (solid), and $M=5$ (dashed-dotted) moment approximations to the
  simulated polydispersity distribution. For comparison, the PY-based
  theoretical result is also shown (cyan line).
  Open circles and triangles mark
  the amplitudes determined from the $\varphi=0.585$
  simulation data via the function
  $Y(q)$, Eq.~\eqref{Yfunc}, and via the von~Schweidler fits discussed
  in conjunction with Fig.~\ref{fqall}, respectively. The results for $Y(q)$
  have been scaled by a factor $0.61$ to account for the unknown normalization
  in this procedure.
}
\label{hq}
\end{figure}

As a cross-check, we have also determined the critical amplitudes
by directly fitting the von~Schweidler expression including its leading-order
correction to the late $\beta$ regime, as described in conjunction with
Eq.~\eqref{vsfits}.
Figure~\ref{hq} shows the results for the critical amplitude of the
coherent density correlators, $h_q$. Reassuringly, both determinations of
$h_q$ give results that are fully consistent with each other. Also shown
in the figure are the MCT-calculated amplitudes. For all three values of
the number of components chosen, $M$,
the data are in very good agreement, with strongest deviations
setting in for $qd\lesssim6$. In particular, the strong dip in $h_q$ around
$q\approx q_p$ is well reproduced. In general, the shape of the $h$-versus-$q$
curve is in this regime quantitatively
captured already by the Percus-Yevick approximation discussed
in Ref.~\cite{Franosch}. At small wave numbers, $qd\lesssim6$,
deviations set in that are the analog of those discussed above in
connection with $f_q^c$ and $\tau_q$: polydispersity affects the
long-wavelength fluctuations in the system, and using the pre-averaged
static structure factor within MCT cannot describe these mixture-specific
features. It is intuitively clear that, since the actual $f_q^c$ is larger
than its one-component estimate at small $qd$, the opposite has to hold
for $h_q$, as the normalization of the correlation function implies
$f_q+h_q<1$.

\begin{figure}
\includegraphics[width=.9\linewidth]{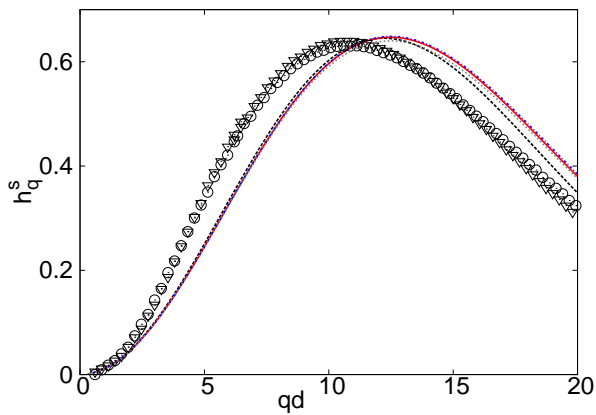}
\caption{\coloronline
  Critical amplitudes $h_q^s$ calculated within MCT for $M=1$ (dashed),
  $M=3$ (solid), and $M=5$ component (dashed-dotted) approximations to the
  polydispersity distribution, and by using the Percus-Yevick static
  structure factor (cyan). Open circles are the corresponding amplitudes
  determined from $Y(q)$, Eq.~\eqref{Yfunc}, scaled by a factor $0.63$.
  Triangles show results from von~Schweidler fits as in Fig.~\ref{fqall}.
}
\label{hqs}
\end{figure}

In Fig.~\ref{hqs}, the tagged-particle critical amplitudes $h^s_q$ are
displayed. Again, the two procedures to determine this quantity from the
simulation data alone agree. The result shows a peak around $qd\approx10$,
while for $q\to0$, $h^s_q\to0$ follows from hydrodynamic laws.
In contrast to the coherent amplitude $h_q$, for $h^s_q$ the MCT results
show more pronounced deviations from the simulation values. The theoretical
quantities exhibit only a weak dependence on the number of component bins
$M$ in this case, and peak around $qd\approx12.5$, i.e., at slightly
larger wave numbers than what is observed in the simulation. Generally,
a shift of the $h^s_q$-versus-$q$ curve arises, indicating that MCT
gives a wrong estimate of the relevant length scale for the tagged-particle
motion in the $\beta$ regime. It is unclear whether this mismatch can
be attributed to the mismatch in critical packing fractions $\varphi^c$
between theory and simulation, as was done for $f^{s,c}_q$. It is also
notable that the disagreement is pronounced only in the tagged-particle
critical amplitude; absorbing it into an effective wave number, as done
in Ref.~\cite{voigtmann2004}, would in fact worsen the agreement for the
collective amplitude as can be seen in Fig.~\ref{hq}.

\section{Full MCT-Analysis}\label{mctanalysis}

Having established the generic MCT scenario for the simulation, we now
present the numerical solutions of the full (non-asymptotic) MCT equations.
In principle, the dynamical correlation functions thus obtained can be
directly compared to the corresponding quantities extracted from the
simulation. However, the mismatch in the $\varphi^c$ values neccessitates
a comparison not at equal densities, but at, in principle, equal separation
from the respective transition points. In the spirit of the MCT asymptotics,
a comparison should involve matching the separation parameter $\sigma$;
however, this is not easily determined for the simulation, since the
true functional dependence between $\sigma$ and the control parameters is
not known. Only asymptotically close to the transition do we have
$\sigma\propto\epsilon$, with a pre-factor that also can only be calculated
within MCT (and thus might be in error). It is therefore practical to
perform a fitting of the packing fractions used in MCT, $\varphi_\text{MCT}$
to the nominal ones used in the simulation, $\varphi$, for each of the
MCT systems with a different number of components $M$. We have performed
this fitting based only on the coherent correlators at $q=q_p$; the comparison
for all other wave numbers, and for all tagged-particle quantities then
is parameter-free.

It should be noted that a similar fitting procedure was already performed
in Ref.~\cite{voigtmann2004} for the tagged-particle data alone. There,
however, it was found that an error in the relevant length scale
(as discussed in connection with Fig.~\ref{hqs}) could be absorbed by
adjusting also the values of $q$ in the comparison. Such a procedure
effectively allows to improve the agreement for the plateau values
in the incoherent correlators, since the $f^{s,c}(q)$ are monotonically
decreasing with increasing $q$. In the present case, no such shifting of
wave numbers is allowed for, since for the collective $f^c(q)$, no such
argument holds.

\subsection{Collective Dynamics}

MCT ascribes the dramatic slowing down in the collective dynamics
approaching a glass transition to a bifurcation scenario, where upon
smooth variations of all control parameters, a qualitative change in the
solutions occurs at long times. In fitting the control parameter
$\varphi_\text{MCT}$, it is therefore essential to check that the relation
$\varphi_\text{MCT}(\varphi)$ does not show signs of singular variation
itself. In the ideal case, this relation should be linear, as long as one
considers density intervals where $\varphi$ is still large compared to the
shift $\Delta\varphi^c=(\varphi^c-\varphi^c_\text{MCT})$. Then, such a
relation indicates that the $\sigma$ values calculated from the theory
agree with the `real' ones describing the simulation. Generally,
$\sigma$ is some function of the control parameters, $\sigma=C[\varphi]$,
and we are looking at the approximate inverse,
$\varphi_\text{MCT}(\varphi)=C_\text{MCT}^{-1}\circ C[\varphi-\Delta\varphi]$.
Ideally, this would mean that $\varphi_\text{MCT}=a\varphi-b$, with
$a=1$. The approximate nature of MCT and the approximate matching of
the particle interactions will induce deviations from this ideal behavior.

\begin{figure}
\includegraphics[width=8cm]{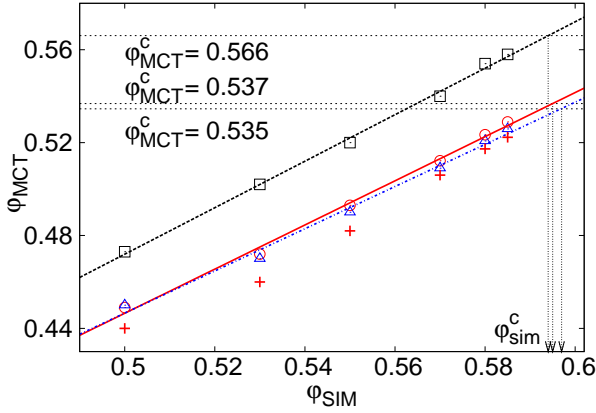}
\caption{\coloronline
  Packing fractions $\varphi_\text{MCT}$ chosen for the five-component
  (triangles), three-component (circles), and one-component (squares) MCT
  fits to the simulation correlators, as functions of the simulated packing
  fraction $\varphi$. The dashed lines are linear regression fits,
  $\varphi_\text{MCT}=a\varphi+b$, with parameters
  $a=1.0001$ ($0.9506$, $0.9083$) and $b=-0.0285$ ($0.0288$, $0.0288$) for
  $M=1$ ($M=3$, $M=5$).
  Crosses mark $\varphi_\text{MCT}$ from a fit to the mean-squared displacement
  only ($M=3$).
  Horizontal dotted lines show the MCT predictions for the critical point
  for each of the three models; their intersections with the linear
  regression curve marks the estimated $\varphi^c$ for the simulation,
  as noted by arrows on the abscissa.}
\label{critpointMD}
\end{figure}

Figure~\ref{critpointMD} shows the resulting $\varphi_\text{MCT}(\varphi)$
for the one-, three- and five-component analyses we performed. We determined this relation by requiring the best possible description of the
species-averaged collective density correlators at $qd=7.3$ (which
essentially involves matching the $\alpha$-relaxation time scale of the
curves).
Linear
regression fits are also shown, and they yield indeed $a\approx1$, so that
the fits we shall discuss below are highly reasonable. Let us point out that
we do not see significant deviations from linearity, not even at the
$\varphi=0.585$ that marks the high-density end of our simulations.
It is usually expected that such deviations set in close to $\varphi^c$
due to the appearance of hopping processes missed in MCT; the commonly reported trends
would appear in Fig.~\ref{critpointMD} as a sublinear growth of the
$\varphi_\text{MCT}$-versus-$\varphi$ curve at high densities, owing to the effects of hopping transport (slower increase in experimental $\tau$ values than predicted by MCT), which would need to be mimicked
in MCT by a saturation in the $\varphi_\text{MCT}$ variation. These
trends are typically reported once the $\alpha$-relaxation time
exceeds the characteristic time of microscopic short-time motion by $10^3$
(the most recent claims are for colloidal hard-sphere like systems
\cite{Brambilla.2009}). 
Our simulations are clearly inside that regime, however we do not find
such deviations. Note however that many previous studies based their
conclusions on relaxation times obtained for tagged-particle quantities,
while we center the discussion on collective density correlators.
Differences may thus partly be attributed to the fact that not all coherent
incoherent relaxation times diverge in the same manner as predicted by MCT.
Indeed, fitting the mean-squared displacement alone, a different
$\varphi_\text{MCT}(\varphi)$ relation is obtained \cite{voigtmann2004},
leading to a different
estimate for the critical point, and different deviations from
linearity. In Fig.~\ref{critpointMD}, crosses mark this relation
for the $M=3$ system. More data points are required to address this
issue further \cite{ley}.

From the curves shown in Fig.~\ref{critpointMD}, one can read off the
estimated critical packing fractions $\varphi^c$ for the simulation data,
as determined from the full MCT analysis once $\varphi^c_\text{MCT}$ is
known (the latter are shown as horizontal dotted lines). We obtain
$\varphi^c\approx0.594$ ($0.595$, $0.597$) for the $M=1$ ($M=3$, $M=5$)
analysis, i.e.\ all values agree within less than $1\%$ and are
in good agreement with the asymptotic analysis of the simulation
data alone \cite{voigtmann2004}, as expected.

\begin{figure}
\includegraphics[width=.9\linewidth]{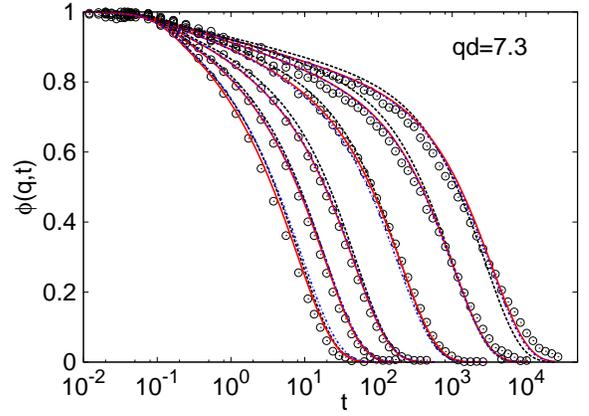}
\caption{\coloronline
  MCT fits of the simulated collective density correlation functions
  (shown as circles) using the simulated static structure factors
  binned into $M=1$ (dashed), $M=3$ (solid), and $M=5$ (dash-dotted lines)
  components to approximate the simulated polydispersity distribution.
  Packing fractions in the simulation are $\varphi=0.5$, $0.53$, $0.55$,
  $0.57$, $0.58$, and $0.585$. The curves have been fitted by adjusting
  only $\varphi_\text{MCT}$ as described in
  conjunction with Fig.~\ref{critpointMD}; we get
  $\varphi_{\text{MCT},M=1}=0.473$, $0.502$, $0.52$, $0.54$, $0.554$, $0.558$;
  $\varphi_{\text{MCT},M=3}=0.449$, $0.472$, $0.493$, $0.5122$, $0.5234$, $0.5289$; and
  $\varphi_{\text{MCT},M=5}=0.45$, $0.47$, $0.49$, $0.509$, $0.5207$, $0.5259$.
}
\label{fitatpeak}
\end{figure}

Figure~\ref{fitatpeak} shows the simulated coherent correlators for
$qd=7.3=q_pd$ together with the MCT curves for the one-component system,
as well as the $M=3$ and $M=5$ systems, for different densities,
from which the relation discussed above in conjunction with
Fig.~\ref{critpointMD} has been fixed.

Although little difference can be seen at this wave-vector magnitude between
the different $M$, the multi-component models give slightly better agreement
with the simulation, mainly because they show a more stretched final decay.
As discussed above, the MCT results yield $\beta_q$ values which are too
high, but the trend with increasing $M$ slightly changes $\beta_q$ in the
right direction. It is nevertheless remarkable, that the shape of the
$\alpha$ relaxation is much better reproduced in the MCT calculation than
one would expect from the difference in $\beta_q$ visible in
Fig.~\ref{betaall} (still about $0.2$ even at $q=q_p$). This clearly indicates
that the Kohlrausch function is at best an approximate characterization of
the $\alpha$-relaxation function at these wave numbers.

Let us remark that the simulation correlators exhibit an unexpected behavior in the very last phase of their decay at high densities e. g. for $\varphi = 0.58$ and $0.585$ at values for $\Phi(q,t)$ below $0.05$. Here they show a strongly decreasing slope (a `foot'). The effects on the Kohlrausch-fits has been determined to be $\pm 3 \%$, by applying the same fit routine and omitting the last part of the curves. Thus this behavior cannot be an explanation for the mismatch in the stretching exponents.

\begin{figure}
\includegraphics[width=.9\linewidth]{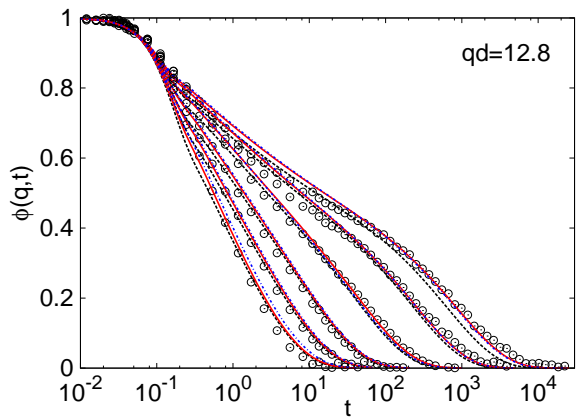}
\caption{\coloronline
  MCT description of the collective density correlation functions from the
  simulation
  at $qd = 12.8$ (corresponding to the second peak of the averaged static
  structure factor). Packing fraction and symbols as in Fig.~\ref{fitatpeak}.
}
\label{fitatsecondpeak}
\end{figure}

Having now fixed all parameters, we compare in Fig.~\ref{fitatsecondpeak}
simulation and MCT correlators for $qd=12.8$, corresponding to the second
peak in $S(q)$. The overall fit quality is found to be the same as for
$q=q_p$, with changes in stretching, relaxation time and plateau height that
are well captured in the theory. Again, curves for different $M$ agree,
with deviations becoming visible at the highest density investigated.
It can also be noted that MCT overestimates the correlators in the
crossover-region from the microscopic to the structural relaxation part.
Such effects will become more apparent; see below.

\begin{figure}
\includegraphics[width=.9\linewidth]{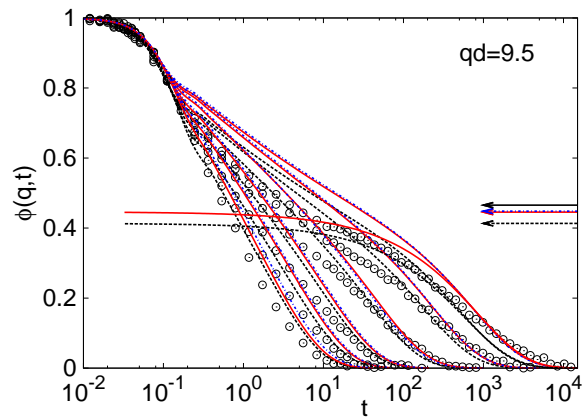}
\caption{\coloronline
  MCT description of collective density correlation functions from the simulation
  at $qd=9.5$ (corresponding to the first dip after the main peak of the
  averaged static structure factor). Packing fraction and symbols as in
  Fig.~\ref{fitatpeak}. 
  Arrows on the right mark the plateau values obtained from Kohlrausch fits
  discussed in Fig.~\ref{fqall}. The two curves starting at the plateau values are the 
  solutions to the corresponding $\alpha$-master equation.}
\label{firstdip}
\end{figure}

The next point of interest is the first dip of the structure factor,
$qd=9.5$. In the region where $S(q)<1$ holds, the differences of the one- and
multi-component MCT calculations are strongest concerning the plateau values
(see Section~\ref{alpha-section}). As a consequence of this, the time-dependent
MCT correlators show stronger discrepancies for one- and multi-component
results. Figure~\ref{firstdip} shows the comparison of these curves with the
simulation data for $qd=9.5$. Indeed, the $M=1$ result clearly deviates
from the $M=3$ and $M=5$ ones. However, none of them gives a satisfying
description of the simulation data in the intermediate time window
$0.1\lesssim t\lesssim\tau_q$; only at the longest times, the MCT
curves describe the data well. This reflects the fact that the
$q$-dependence of $\tau_q$ is well reproduced, see Fig.~\ref{tauall},
so that fitting $\tau$ for $q=q_p$ also gives good agreement for the
final relaxation time at other $q$.

There is an interesting feature observed in Fig.~\ref{firstdip}: while
judging from Fig.~\ref{fqall} the plateau values are in good agreement
between simulation and MCT for $M\ge3$, this agreement is not obvious
in the correlators. The reason is that in the MCT curves, a pronounced
stretched decay of the correlators from about $\phi\approx0.8$ down to
$\phi\approx f$ is visible. The simulation curves in contrast show a clear
shoulder at $\phi\approx f$, following a relatively steep decay from
the short-time regime. The latter is also well known from MD simulations
\cite{kob2003}. It is one of the main problems MCT has in
describing the early $\beta$-relaxation window.

\begin{figure}
\includegraphics[width=.9\linewidth]{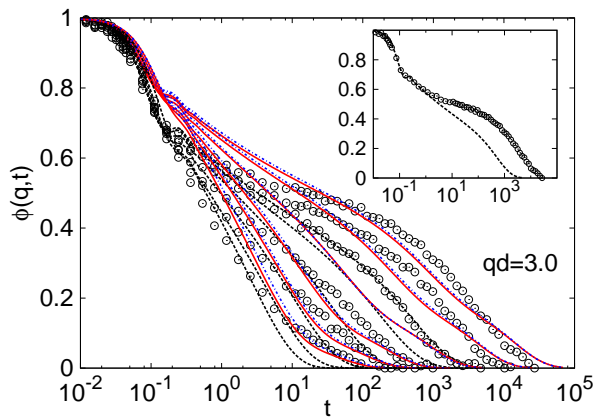}
\caption{\coloronline
  MCT description of collective density correlation functions from the simulation
  at $qd=3.0$, with symbols and packing fractions as in Fig.~\ref{fitatpeak}.
  The inset shows only the highest packing fraction, $\varphi=0.585$, and its corresponding $M=1$ correlator,
  in order to highlight the different relaxation times of the different
  MCT approaches.
}
\label{verylowq}
\end{figure}

In general the behavior seen in Figures \ref{firstdip} and
\ref{fitatsecondpeak} is exemplary: MCT solutions for the $q$-values belonging
to regions where $S(q) < 1$ give worse agreement than the ones belonging to
$q$-values where $S(q)>1$ holds. The root of this problem might be buried in
the short-time relaxation part which lasts for longer times at these
$q$-values and thus could still influence the $\beta$-process.
In one-component systems, the short-time relaxation is given by
$\Phi(q,t)\propto\exp[-q^2/S_q t]$ for a colloidal system. Hence it is
conceivable that for lower $S_q$, the short-time relaxation is slowed down
and hence plays a more important role in the dynamics. Since MCT in
general only rather crudely includes the short-time relaxation,
this might be the cause of the worse agreement.

As expected from Fig.~\ref{fqall} the multi-component approach matches the
simulation correlators much better for low $q$ values. This is exemplified
in Fig.~\ref{verylowq}, where data for $qd=3.0$ are shown. The one-component
MCT solution underestimates the structural relaxation times by one to two
decades (cf.\ the inset of Fig.~\ref{verylowq}),
while already $M=3$ gives an overall fit to the data that is much
better. This is of course another manifestation of the qualitative change
in $\tau_q$ for small $q$ discussed in conjunction with Fig.~\ref{tauall},
and hence a signature of the interdiffusion process that is absent in
the $M=1$ calculation.

Especially at the highest densities, one notices in the MCT solutions
for $M=3$ and $M=5$ the emergence of a double $\alpha$-relaxation phenomenon,
visible as a shoulder around $\phi\approx0.2$. Remarkably, both the
$M=3$ and $M=5$ results are in close agreement.
The simulation data does
not show such a double-relaxation pattern, which we attribute to the fact
that the simulation is truly polydisperse (containing as many species
as there are particles), and that thus the different $\alpha$ relaxations
stemming from the superposition of a structural relaxation with
different interdiffusion processes are smeared out.

At lower densities, the signature of the interdiffusion process remains
in the MCT curves as a kink in the relaxation curve for $\phi\approx0.1$,
followed by a `foot' that extends almost over one decade in time at
the lowest $\varphi$ shown in Fig.~\ref{verylowq}. Interestingly, the
simulation data for this density also show such a feature, and are in fact,
apart from a shift in time scale for the short-time motion, well described
by the MCT curve for $M\ge3$. It will be worthwhile checking for the
generality of such a `foot' in the relaxation functions of glass forming
systems (as most of them are mixtures).

\subsection{Tagged-Particle Dynamics}

\begin{figure}
\includegraphics[width=.9\linewidth]{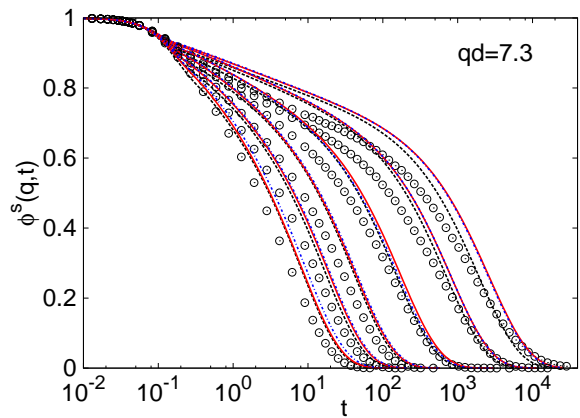}
\caption{\coloronline
  MCT description of the tagged-particle density correlation functions
  determined from the simulation at $qd=7.3$; packing fractions and symbols as
  in Fig.~\ref{fitatpeak}.
}
\label{incoherent}
\end{figure}

We now address the quality of the MCT description for the tagged-particle
correlation functions, after all adjustable parameters have been fixed
through an analysis of the collective density fluctuations at $q_p$.
Figure~\ref{incoherent} shows the results for $\phi^s(q,t)$ at
$q=q_p = 7.3$,
i.e., it is to be compared to Fig.~\ref{fitatpeak}.
Two trends are visible in Fig.~\ref{incoherent} that mark the main
shortcomings of MCT in describing the tagged-particle dynamics: first,
the relaxation to the plateau values is too slow, like in Fig.~\ref{firstdip}.
Secondly, a shift in the $\alpha$-relaxation time is noted, in agreement
with the expectation from Fig.~\ref{tauall}: the MCT curves decay too
slowly, although the collective relaxation times 
match those of the simulation at the same wave number.
Another general finding for the tagged-particle dynamics is that the
number of component bins $M$ to model the polydisperse distribution has
little influence on the quality of the description.

\begin{figure}
\includegraphics[width=.9\linewidth]{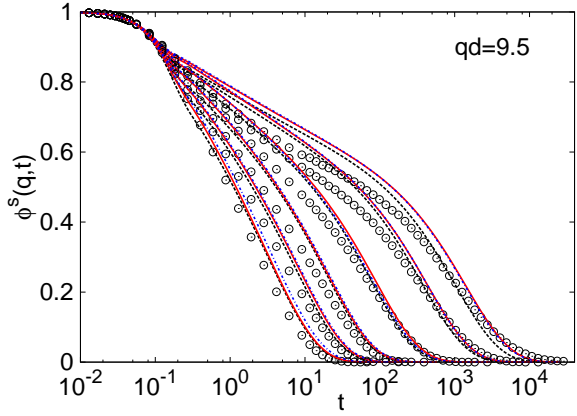}
\caption{\coloronline
  MCT description of the tagged-particle density correlation functions
  as in Fig.~\ref{incoherent}, but for $qd=9.5$.
}
\label{incoherent_9.5}
\end{figure}

The agreement in $\tau^s_q$ improves somewhat with higher $q$;
in agreement with this, also the tagged-particle correlators are somewhat
better described by MCT for larger $q$, as shown for the exemplary case
$qd=9.5$ in Fig.~\ref{incoherent_9.5}. However, the mismatch in the plateau
region remains roughly the same as in Fig.~\ref{incoherent}.

\begin{figure}
\includegraphics[width=8cm]{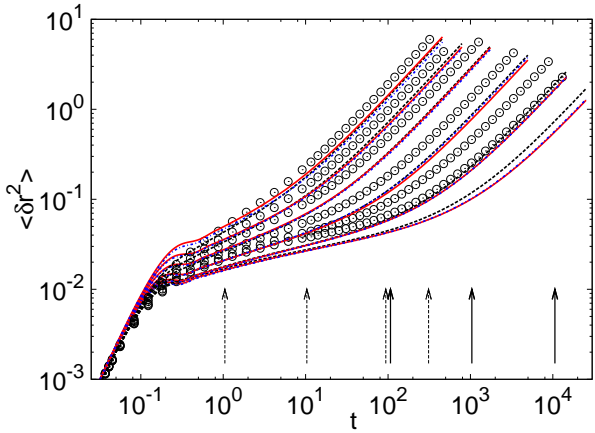}
\caption{\coloronline
  Mean squared displacement from the simulation (open circles) compared with
  their MCT description based on the fit performed in Fig.~\ref{fitatpeak};
  different line styles correspond to $M=1$, $M=3$, and $M=5$ models of the
  polydispersity distribution, cf.\ Fig.~\ref{fitatpeak}.
  Arrows indicate the times for which the van~Hove functions are shown
  in Fig.~\ref{vHove_log_lowpack} (dashed) and Fig.~\ref{vHove_log}
  (solid arrows).
}
\label{msd}
\end{figure}

The worsening of the quality of the MCT fits with decreasing $q$ for the
tagged-particle quantities was already noted in Ref.~\cite{voigtmann2004},
and seems to point to an inherent error in the theory, that is, however,
too poorly understood to be improved upon. The error furthermore increases
with increasing packing fraction, and cannot be eliminated by alluding
to polydispersity effects. It is, of course, directly connected to the
well known decoupling between diffusivity and finite-$q$ local
relaxation times, upon which we will embark again below.
It is therefore not surprising that the most
drastic deviations between simulation and MCT are visible in the
mean-squared displacements, shown in Fig.~\ref{msd}, since this corresponds
to the $q\to0$ limit of tagged-particle density correlations. The
long-time relaxation leading to the final diffusive part in the MSD
is much slower in MCT than it is in the simulations; while at
$\varphi=0.5$, for the long time diffusion 
both curves agree within $24 \%$, at $\varphi=0.585$,
the discrepancy is about  a factor $3$.
The agreement is again not remedied by including a better description
of polydispersity effects; in fact the agreement is somewhat worsened
in the $M=3$ and $M=5$ calculations at $\varphi=0.585$ as compared to the
one-component analysis. Only the shape of the MSD is well described by
MCT, again with the caveat that the relevant length scale, in this case
the height of the plateau (indicating the squared cage size up to a
trivial prefactor), is in error in MCT; the theory underestimates the
localization length by about $10\%$. This quantitative error agrees with
the one found for the tagged-particle critical amplitude $h^s_q$ in
Fig.~\ref{hqs} (and cannot be accounted for by recalling that the
MCT-calculated critical density is too small).
Fitting only the MSD by adjusting $\varphi_\text{MCT}(\varphi)$
based on this data 
alone, the simulation data can be described quantitatively (see
Ref.~\cite{voigtmann2004} and the discussion of Fig.~\ref{critpointMD}).
Similar conclusions have been drawn from an analysis of
dynamic-light-scattering experiments on colloidal suspensions
\cite{sperlmsd,vanmegen2007b}.

Following an approach suggested in Refs.~\cite{cates2004, puertas2004, reichmann2005,FlennerSzamel2005,FlennerSzamel2005_2} 
we investigated the probability distributions of the logarithm of
single-particle displacements $P(\log_{10} (\delta r),t)$ at a time $t$. The
appearance of different peaks in $P(\log_{10} (\delta r),t)$ is a result of
populations of particles with different mobilities, and was suggested as origin of
the failure of MCT to capture the dynamics of the MSD \cite{FlennerSzamel2005,FlennerSzamel2005_2}.
The probability distribution is directly related to
the van~Hove function via $P(\log_{10} (\delta r),t) = \ln (10) 4 \pi \delta
r^3 G_s(\delta r, t)$, and its shape
is independent of time for a Gaussian van~Hove function
$G_s(\delta r, t) = 1/(4 \pi Dt)^{3/2} \exp (- \delta r^2 /4Dt)$
\cite{FlennerSzamel2005_2}.

\begin{figure}
\includegraphics[width=.9\linewidth]{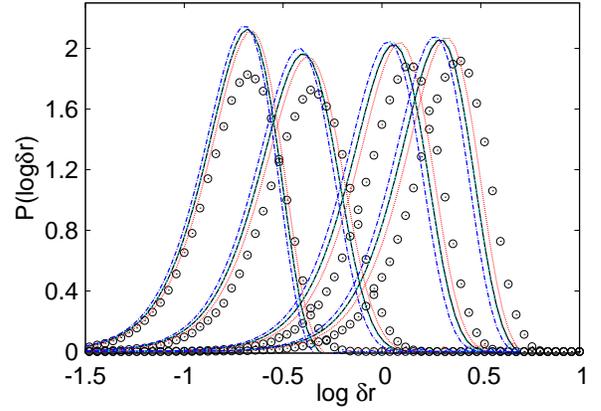}
\caption{\coloronline
  $P(\log \delta r) = \ln(10) 4 \pi \delta r^3 G_s(r,t)$ for the times marked
  by the dashed arrows in Fig.~\ref{msd}. Open circles are the simulation
  results for $\varphi=0.5$, while lines indicate the corresponding MCT result
  for the $M=3$ multi-component model at $\varphi_\text{MCT}=0.449$:
  solid black lines denote the full distribution, while red-dashed (cyan-short-dashed,
  blue-dashed-dotted) mark the distributions involving only small (medium, large)
  particles.
}
\label{vHove_log_lowpack}
\end{figure}

Figure~\ref{vHove_log_lowpack} shows $P(\log_{10} (\delta r),t) $ for the
lowest packing fraction, $\varphi=0.5$, and its MCT fits by the $M=3$
model, ($\varphi_\text{MCT}=0.449$).
Both simulation and MCT show van~Hove functions that deviate little from
the Gaussian expected for ordinary diffusion. As already clear from the
mean-squared displacements, Fig.~\ref{msd}, the peak position in MCT is
at lower $\delta r$ values than in the simulation.
The MCT calculation is also shown separated into the three different
particle sizes; quite intuitively, the smaller particles are predicted
to move faster on average, hence the peak visible in Fig.~\ref{vHove_log_lowpack}
shifts to the right when considering only the small particles. One can
imagine that the mean-squared displacement is a quantity that is dominated
by motion of fast particles, so that an improvement on the theoretical
result may be to give stronger weight to their displacements.
The resulting shift
is however seen from Fig.~\ref{vHove_log_lowpack} to be insufficient to
quantitatively explain the difference to
the simulation data.

\begin{figure}
\includegraphics[width=.9\linewidth]{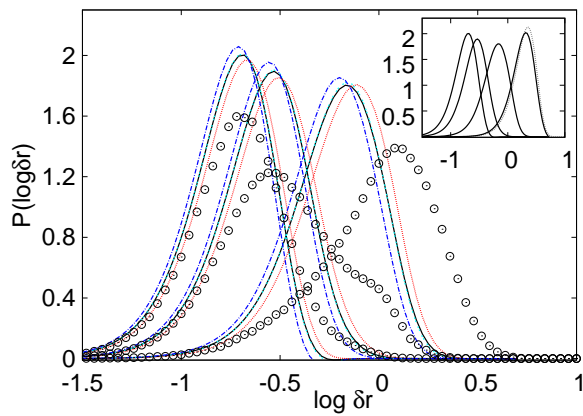}
\caption{\coloronline
  $P(\log \delta r)$ as in Fig.~\ref{vHove_log_lowpack}, for the times
  marked by solid arrows in Fig.~\ref{msd} and for packing fraction
  $\varphi=0.585$ (fitted by $\varphi_{\text{MCT},M=3}=0.5289$).
  The inset shows the average MCT distributions with an additional result
  at $t=1.01 \; 10^{5}$ where long-time diffusion has already set in.
  The dotted grey line is a fit with a Gaussian distribution,
  $(4\pi Dt)^{-3/2}\exp[-\delta r^2/(4Dt)]$, where
  $D=8.35\times10^{-6}$ is taken from Fig.~\ref{msd}.}
\label{vHove_log}
\end{figure}

At higher packing fractions, the MCT description worsens still, and one starts
to see
in the simulation strong deviations from Gaussian behavior. This is
shown in Fig.~\ref{vHove_log} for the case $\varphi=0.585$
($\varphi_\text{MCT}=0.5289$). A second shoulder in the distribution
$P(\log_{10}(\delta r),t)$ at intermediate times can be interpreted as
``hopping-like'' motion for a certain fraction of particles \cite{FlennerSzamel2005}.
Such emergent two-peak structures are also known from colloidal gels
\cite{puertas2004}, and binary Lennard Jones mixtures \cite{FlennerSzamel2005, FlennerSzamel2005_2, szamel2006_2}.

\begin{figure}
\includegraphics[width=8cm]{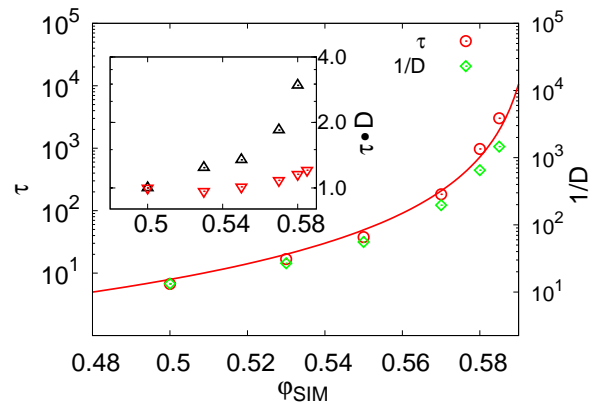}
\caption{\coloronline
  $\alpha$-relaxation time scale $\tau$ determined from the master curve
  of Fig.~\ref{rescaled-cor} (red circles), and inverse long-time
  self-diffusion coefficients $1/D$ extracted from the simulated
  mean-squared displacements (green diamonds). The red-solid line shows the
  power law $\tau \sim|\epsilon|^{-\gamma}$ with $\gamma=2.445$ determined
  from the $M=3$ MCT calculations.
  Inset: product $D\tau$ evaluated from the simulation data (black up-triangles)
  and from MCT (red down-triangles) scaled by a factor to make them comparable. 
}
\label{powerlaw}
\end{figure}

The appearance of dynamical heterogeneities as signalled by
Fig.~\ref{vHove_log} is usually connected with the decoupling of diffusive
and collective (viscous) time scales, i.e., the breakdown of the
Stokes-Einstein (SE) relation mentioned in the introduction. MCT predicts the
SE relation to hold  close to $\varphi^c$, as both the
inverse of the long-time self-diffusion coefficient, $1/D$, and the
typical $\alpha$-relaxation time scale $\tau_q$ should diverge with the
same asymptotic power law, so that $D\tau_q$ approaches a constant as
$\varphi\to\varphi^c$ from below. Figure~\ref{powerlaw} displays a typical
collective relaxation time, $\tau$, i.e., the one extracted from the
determination of the $\alpha$-master curves, Fig.~\ref{rescaled-cor},
and the inverse diffusion coefficient as functions of $\varphi$ for the
simulation results.
Over the window accessible in our simulations, both quantities do show
power-law-like divergence, and in particular $\tau$ can be reasonably well
fitted with the expected MCT asymptote, $\tau \sim|\sigma|^{-\gamma}$,
assuming that $\epsilon\propto\sigma$ and $\gamma=2.445$ can be taken from
the $M=3$ calculation.
The von~Schweidler fits in Fig.~\ref{kohl} correspond to $\gamma \approx 2.63$, so that the MCT exponent relations are (alomost) fulfilled. The diffusivities, however, are better described by a similar power law
with exponent $\gamma_\text{MSD}=2.07942$.

The decoupling of viscous and diffusive time scales is best exhibited
by taking the product $D\tau$, which is shown in the inset of
Fig.~\ref{powerlaw}. One clearly notes in the simulation data a change
of about a factor $4$, and no tendency to approach a constant at the
highest $\varphi$ we can investigate.

\begin{figure}
\includegraphics[width=8cm]{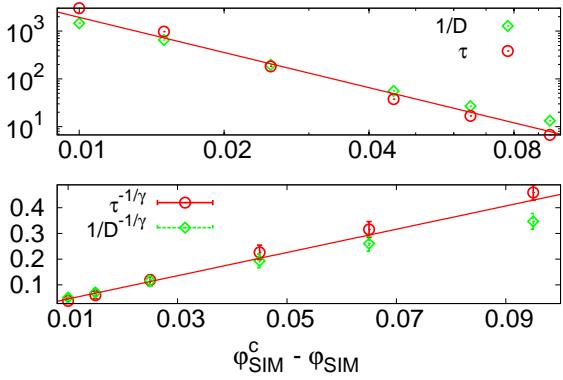}
\caption{\coloronline Upper panel: Inverse self Diffusion coefficient $1/D$ (green diamonds) and $\alpha$-relaxation time scale $\tau$ determined from the master curve of Fig.~\ref{rescaled-cor} (red circles) with double logarithmic axes. Lower panel:  ${1/D}^{-1/\gamma}$ (green diamonds) and $\tau^{-1/\gamma}$ (red circles) from the upper panel. The errobars are estimated from the different $\gamma$-values obtained from the MCT calculations for $M=1,3,5$ components. In both panels the solid red line shows the corresponding power law from Fig.~\ref{powerlaw}.
}
\label{rectification}
\end{figure}

Rectification plots as shown in Fig.~\ref{rectification} corroborate that the asymtotic powerlaw holds for the collective relaxation time $\tau$ over at least two decades, and that a differing exponent or a different critical density would be required to render the diffusivity as a compatible MCT power law.

\begin{figure}
\includegraphics[width=\linewidth]{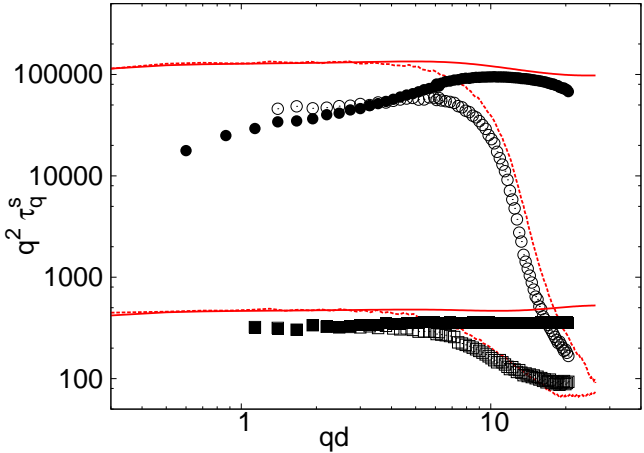}
\caption{\coloronline
  Relaxation times $\tau^s_q$ of the tagged-particle density correlation
  functions at packing fraction $\varphi=0.585$ (upper panel;
  $\varphi_\text{MCT}=0.5289$) and
  $\varphi=0.5$ (lower panel; $\varphi_\text{MCT}=0.449$),
  plotted as $q^2\tau^s_q$. Filled
  symbols (full lines): $\tau^s_q$ extracted from Kohlrausch fits to the
  simulation (theory) data, taking
  into account the plateau values $f^s_q$. Open symbols (dotted lines):
  determined via $\phi^s(q,\tau^s_q)=1/e$.
}
\label{szamelfig}
\end{figure}

It is instructive to compare the wave-vector dependence of the relaxation times
$\tau^s(q)$ for the tagged-particle correlation function with the limiting
behavior expected on hydrodynamic grounds, $\tau^s(q\to0)\sim1/q^2D$.
To this end, we plot in Fig.~\ref{szamelfig} $q^2\tau^s(q)$ extracted from
both the simulation and our MCT fits, at representative low and high
packing fractions. Such plots have been suggested in discussing the
non-Fickian transport evidenced by the van~Hove functions shown above
\cite{Berthier,FlennerSzamel2005_2}. There, $\tau^s(q)$ has been defined as the
point where the $q$-dependent tagged-particle correlation function has
decayed to $1/e$. This quantity (open symbols and dashed lines
in Fig.~\ref{szamelfig}) approaches $1/D$ at low $q$. At large $q$
it is expected to drop sharply: as the amplitude of the structural relaxation
process, $f^s_q$, drops below $1/e$, the procedure no longer reliably
probes slow relaxation, but rather is dominated by
the microscopic short-time relaxation (essentially $1/D_0$ in a Brownian
system, where $1/D_0\ll1/D$). A similar remark holds for the collective
relaxation time $\tau(q)$; the time-scale determined by the
$1/e$-criterion can only be compared to MCT as long as $f_q$ is sufficiently
larger than $1/e$.

Our findings for $\tau^s(q)$
are in agreement with the Brownian dynamics simulations of
Flenner and Szamel \cite{FlennerSzamel2005_2}: while MCT simply predicts a
monotonic crossover between the two regimes, in the simulation data, an
intermediate maximum at $qd$ corresponding to the nearest-neighbor
distance emerges as one approaches $\varphi^c$.

Since our MCT fits are matched to the \emph{collective} correlation functions,
a notable consequence in Fig.~\ref{szamelfig} is that for $q\to0$, MCT
and simulation data for $q^2\tau^s(q)$ approach different constants, the
MCT one being too high. This corroborates our interpretation that for the
tagged-particle dynamics MCT
is a reasonable theory for the intermediate and large wave numbers,
and that deviations are increasingly seen as $q$ approaches zero.
It differs from the interpretation
of Ref.~\cite{FlennerSzamel2005_2}, where also simulation and MCT data for
$q^2\tau^s(q)$ were compared, but normalized to their respective $q\to0$
values. As a result, deviations were attributed mostly to the intermediate-$q$,
not the small-$q$ regime. In light of our results, it might be more
suggestive to turn around the discussion: it is not the nearest-neighbor-scale
modes that are unexpectedly slow, but it is the diffusion that is faster
than expected from the MCT-embedded cage picture.

\begin{figure}
\includegraphics[width=\linewidth]{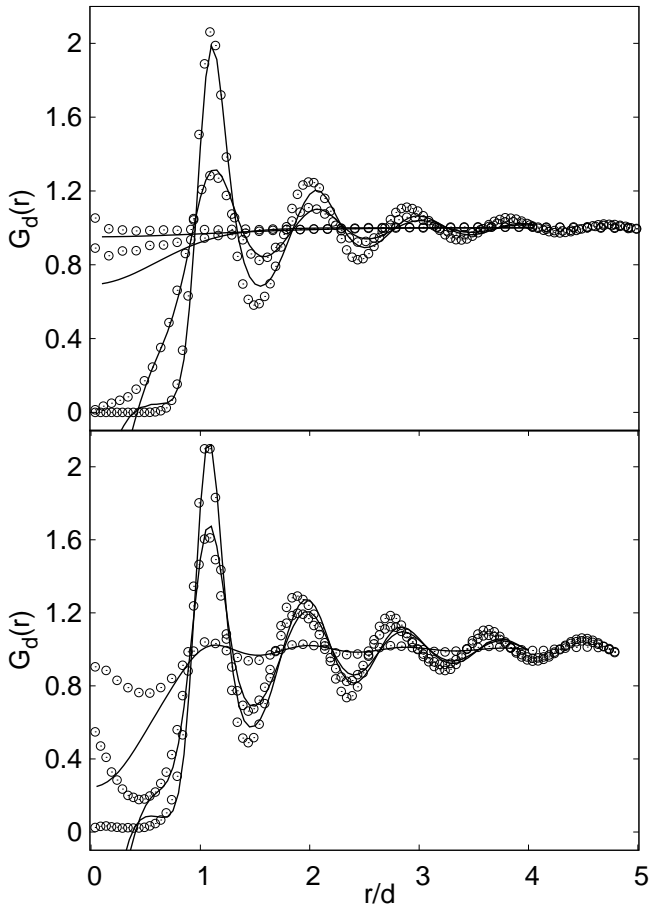}
\caption{
  Distinct van~Hove correlation fundtions $G_d(r)$ at packing fractions
  $\varphi=0.5$ (upper panel) and $0.585$ (lower panel) as obtained by
  simulation (symbols), at various times as indicated by arrows in
  Fig.~\protect\ref{msd}. Lines are the
  corresponding MCT fits evaluated by inverse Fourier transform of
  the difference between collective and tagged-particle density
  correlation functions.
}
\label{vanhovecoh}
\end{figure}

If one investigates the distinct part of the collective dynamics, i.e.,
density correlations that arise from distinct particles, MCT's
mis-description of tagged-particle dynamics has an interesting consequence.
Recall that the distinct van~Hove correlation function $G_d(r,t)$ is given by
the
difference of collective and tagged-particle contributions,
$G_d(r)=G(r)-G_s(r)$; this means that within MCT, it is obtained from the
inverse Fourier transform of $\Phi(q,t)-\phi^s(q,t)$.
On physical grounds, $G_d(r)$ must be a positive real function,
since it measures the probability of finding a particle at time $t$ and
distance $r$ from a distinct particle that was at the origin at $t=0$.
This property is not obvious from the difference formula.

Symbols in Fig.~\ref{vanhovecoh} show $G_d(r,t)$ evaluated at various times
covering the structural-relaxation regime for the simulations at the
lowest and highest packing fraction we studied; we have normalized $G_d(r)$
such that it approaches unity at long distances. We recover the expected
shell structure that is inherited from the $n$-th neighbor shells in the
radial distribution function $g(r)=G_d(r,t=0)$. These shells are increasingly
washed out as time progresses, until the long-time limit $G_d(r,t=\infty)=1$
is reached. Comparing with the MCT-calculated quantities (obtained from the
fits in $q$-space presented above), we note that at distances $r$ including
and exceeding the nearest-neighbor distance, the MCT description is fairly
good although not perfect. At small $r$, however, there is a most obvious
error as $G_d(r,t)$ turns negative in the MCT approximation. Note that for
small and for sufficiently large $t$, this phenomenon does not occur, for
trivial reasons; for small $t$, the positiveness of $G_d(r,t)$ within MCT
hinges on that of $g(r)$, while for large $t$, $G_d(r,t)$ approaches the
uniform density. Note that the positivity of $g(r)$ may fail for approximate
$S(q)$ at some densities, but we have checked that this is not the case here.

The reason for the failure in the small-$r$ description is easily understood:
both $G(r,t)$ and $G_s(r,t)$ are dominated by a strong peak centered on
$r=0$, since their $t=0$ values incorporate a $\delta$-peak that is
smeared out with time. Evaluating $G_d(r,t)$, we have to subtract these
two large contributions from each other. In fact, for the plots shown,
typical values of $G(r=0,t)$ and $G_s(r=0,t)$ are $\approx14$ at the
intermediate time, while
$|G_d(r=0,t)|/G(r=0,t)={\mathcal O}(0.1)$.
Numerically, this is a moderate error, which
by looking at the distinct van~Hove function is turned into a qualitative one.
The error was in fact to be expected based on our discussion so far:
the theory, by way of underestimating the single-particle diffusion
coefficient, overestimates the localization of a tagged particle.
This translates into a peak in $G_s(r,t)$ that is to narrow and thus too
high. Even if the description of the collective dynamics through
$G(r,t)$ were totally correct, this overestimation of single-particle
localization is sufficient to render $G_d(r,t)$ unphysically negative.

In the simulation data, one notices a subtle feature around $r=0$ that
is, due to the reasons just outlined, outside the scope of MCT.
At $r<d$, $G_d(r,t)$ raises from
zero at short times to unity at long times.
At the low density shown in Fig.~\ref{vanhovecoh}, this filling in of the
excluded-volume gap happens as na\"\i{}vely expected from the broadening
of the nearest-neighbor peak, resulting in functions $G_d(r,t)$ that are
always monotonically increasing with $r$ in the regime $r<d$. At higher
density, however, this monotonicity is lost, and an additional dip in
$G_d(r,t)$ evolves around $r=d/2$. This qualitatively agrees with earlier
findings from simulations of glass-forming binary mixtures
\cite{KA,SS}. It is intuitively interpreted as
the persistence of preferred interparticle distances: as a given particle
moves away from its original position, there is an enhanced probability
that \emph{another} particle fills this position, rather than any nearby
one.

\begin{figure}
\includegraphics[width=\linewidth]{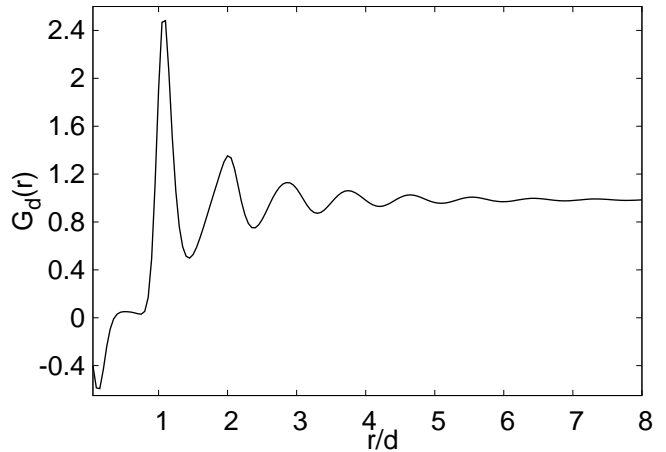}
\caption{
  Distinct part of the van~Hove correlation function $G_d(r)$ that
  corresponds to the MCT plateau values of a monodisperse hard-sphere
  system within the Percus-Yevick approximation, as a function of
  distance in units of the sphere diameters.  
}
\label{vanhovecoh_fr}
\end{figure}

It is worth noting that the issue of negative $G_d(r)$ does not
appear to be related to the dynamics. As shown in Fig.~\ref{vanhovecoh_fr},
the real-space representation of the nonergodicity parameters, calculated
from the inverse Fourier transform of $S(q)f(q)-f^s(q)$, shows the same
features as discussed above. To ensure that no cutoff problems arise,
we have based this quantity on the Percus-Yevick static structure factor
with a wave-vector grid spacing $\Delta q=0.05$ and $M=1600$ discretization
points, and at packing fraction $\varphi=0.516$ for one-component MCT.
Note that we are still sufficiently far from the packing fraction
$\varphi\ge0.6$ where the PY-$g(r)$ starts to be unphysical.

\section{Conclusion}\label{concl}

We performed molecular-dynamics computer simulations of a polydisperse
quasi-hard sphere system and analysed both the collective and the incoherent
density-density correlation functions in the framework of both asymptotic
predictions and full numerical solutions of the mode-coupling theory
of the glass transition. For the latter, the required input in the form
of static equilibrium structure factors has been also calculated from the
MD simulation. To capture some essential effects of particle-size
polydispersity in the MCT calculation, numerical solutions of one-, three-,
and five-component systems with particle sizes chosen to match the
first few moments of the true polydispersity distribution have been
compared.

For the particular size distribution chosen in the simulation, the
five-component analysis turned out to be mostly sufficient to capture
the effects induced by a variation of particle size. These effects concern
in particular the small-wave-number limit, $q\to0$, where mixtures of
particles of unequal interactions (here: unequal sizes) show, even in
the species-averaged correlation functions, signatures of interdiffusion
processes that are well understood in principle in the framework of
multicomponent hydrodynamics. MCT has this hydrodynamic limit built in,
and consequently predicts a subtle interplay of the various interdiffusion
modes with structural relaxation, leading at low $q$ to the appearance
of double-$\alpha$ peaks. These are less pronounced in the simulation,
presumably due to them being smeared out in a truly polydisperse system,
where the binning into a few number of components is not strictly meaningful.

Apart from the low-$q$ behavior, which however may be of significant
interest as it determines the most commonly discussed transport coefficients,
there is little difference between the different mixtures we considered.
A general trend is that, taking the species average already on the level
of static structure correlations, before entering MCT, is worse than
performing this average only on the dynamic level. This is intuitively
understood from the fact that such pre-averaging tends to smear out the
correlations visible in the static structure factor as oscillations at
large $q$. For systems where the slow dynamics is driven by excluded-volume
interactions, such as our quasi-hard-sphere system, this is a relatively
minor quantitative effect. For systems where short-range interactions
are crucial for the slow dynamics
(affecting the large-$q$ tail of the structure functions), it will be
even more crucial to treat polydisperse systems as mixtures rather than
as effective-one-component systems.

Based on the simulation data and the known asymptotic results from MCT alone,
we have first performed a traditional scaling analysis, revealing the
coefficients involved in desribing the slow $\alpha$ relaxation, and the
$\beta$-relaxation window at intermediate times. The coefficients, such
as the plateau heights and critical amplitudes show good quantitative
agreement with the corresponding quantities calculated within full MCT.
A stretched-exponential relaxation analysis for the $\alpha$ process
reveals some mismatch in the stretching indices.

In fitting the full numerical MCT solutions to the data, only the relation
between $\varphi_\text{MCT}$, the packing fraction entering the MCT
vertex, and $\varphi$, the nominal packing fraction in the simulation,
was adjusted, to absorb the well-known error in the numerical value
of $\varphi^c$ when calculated within full MCT. Indeed, the resulting
relation can be very well described as linear with slope unity, so that
effectively no fit parameter remains (or just one, if one accounts for
the slope being slightly different from unity).

Basing this adjustment on the collective correlation functions for a single
wave vector magnitude, $qd=7.3$, we obtain good agreement between MCT and
the simulation data for the collective density correlators at essentially
all sufficiently large $qd$. Deviations are most prominent where $S(q)<1$.
At small $qd$, the one-component analysis shows severe deviations,
in particular an order-of-magnitude mismatch in the relaxation time,
attributed to the missing interdiffusion process discussed above. The
three- and five-component MCT calculations do not suffer from this
shortcoming and describe also the collective small-$q$ behavior reasonably
well.

We have discussed the MCT description of the tagged-particle dynamics
after all parameters have been fixed in the analysis of collective
correlation functions. Here, the situation is more subtle: for $qd$ exceeding
roughly half the position of the structure-factor main peak, the
MCT description is again reasonable. However, for $q\to0$, errors
increase continuously, to become most prominent in an analysis of the
mean-squared displacement (as has been noted in an earlier publication).
Here, the MCT curves show a much stronger slowing down with increasing
packing fraction, while the simulations exhibit averaged particle mobilities
that are higher than those expected from either MCT or a Stokes-Einstein
argument. This diffusion-relaxation decoupling is of course well known
in the glass literature.
The single-particle motion has been thoroughly investigated before in
terms of van~Hove functions, identifying subsets of fast and slow particles
even in one-component systems \cite{kumar_szamel}.
Also in our system, such a splitting is observed. 
However, we wish to stress that it seems to
affect mostly the MCT description of tagged-particle dynamics. At the same
time, the MCT description of the collective density fluctuations remains
remarkably accurate, and in particular the structural relaxation time
extracted from the simulation does not show any significant deviation
from the values predicted by the theory.
For future work an analysis of the collective van~Hove functions might 
provide additional information about the origin of these deviations. 

We can of course not exclude the possibility that at even higher densities
than those we could simulate, such deviations eventually set in. However,
if this is the case, they need not coincide with the features typically
discussed in terms of heterogeneous dynamics, viz.\ the decoupling of
diffusivity from structural relaxation. It may be that our choice of a
system driven by stochastic dynamics and without a significant energy
scale in the particle interactions is fortuitous. However, this remains to
be clarified. Within MCT, the
independence of structural-relaxation properties of the time-evolution
operator is a major result, confirmed before \cite{gleim2000}. There are
arguments that this correspondence does indeed extend also to the way
simulation results deviate from MCT, but not to higher-order correlation
functions such as four-point susceptibilities \cite{szamel2006,berthier_kob}.

It appears then that fitting incoherent correlation functions with MCT
is a rather roundabout way of testing the theory, and that in particular
it represents an unfortunate test (for the theory, at least) in that
these quantities show strongest deviations from the predicted behavior.
Unfortunately, the MSD is perhaps the quantity most often analyzed in
simulation and in assessing MCT's validity (since it is easy to compute
with good statistics). Our analysis shows that it is the least valuable
quantity to compare MCT with.

Our discussion suggests that other collective correlation functions
provide a fruitful basis for further investigations of MCT and its
approximations: in particular stress-stress auto-correlation functions
should be analyzed, as they are cloesly linked to the memory kernel of
the theory. However, these require even more computational effort to
determine from simulation than the collective density correlators we have
examined here: for the latter, we have averaged $1000$ independent
evaluations, while for the former, up to $4000$ had to be used in previous
work on a similar system with Newtonian dynamics \cite{puertasjcp}.

\begin{acknowledgments}
A.M.P. acknowledges financial support from the Spanish M.E.C. -- project MAT2009-14234-CO3-02.
Th.V. holds a Helmholtz-University Young Researcher Group fellowship (HGF VH-NG406), and is fellow of the Zukunftskolleg of the Universit\"at Konstanz.
\end{acknowledgments}

\bibliography{lit}
\bibliographystyle{apsrev}

\end{document}